\documentclass[iop]{emulateapj}
\usepackage{amsmath}
\usepackage{natbib}
\usepackage{psfrag}
\usepackage{multirow}
\usepackage{hyperref}
\begin{document}
\title{Assessing the influence of the solar orbit on terrestrial biodiversity}
\author{F.\ Feng and C.A.L.\ Bailer-Jones}
\affil{Max Planck Institute for Astronomy, D-69117 Heidelberg, Germany}
\begin{abstract} The terrestrial fossil record shows a significant variation in the extinction and origination rates of species during the past half billion years. Numerous studies have claimed an association between this variation and the motion of the Sun around the Galaxy, invoking the modulation of cosmic rays, gamma rays and comet impact frequency as a cause of this biodiversity variation. However, some of these studies exhibit methodological problems, or were based on coarse assumptions (such as a strict periodicity of the solar orbit).  Here we investigate this link in more detail, using a model of the Galaxy to reconstruct the solar orbit and thus a predictive model of the temporal variation of the extinction rate due to astronomical mechanisms.  We compare these predictions as well as those of various reference models with paleontological data.  Our approach involves Bayesian model comparison, which takes into account the uncertainties in the paleontological data as well as the distribution of solar orbits consistent with the uncertainties in the astronomical data.  We find that various versions of the orbital model are not favored beyond simpler reference models.  In particular, the distribution of mass extinction events can be explained just as well by a uniform random distribution as by any other model tested.  Although our negative results on the orbital model are robust to changes in the Galaxy model, the Sun's coordinates and the errors in the data, we also find that it would be very difficult to positively identify the orbital model even if it were the true one. (In contrast, we do find evidence against simpler periodic models.)  Thus while we cannot rule out there being some connection between solar motion and biodiversity variations on the Earth, we conclude that it is difficult to give convincing positive conclusions of such a connection using current data.  
\end{abstract}


\section{Introduction}\label{sec:introduction}

\subsection{Background}

Over the course of Earth's history, evolution has produced a wide variety of life. This is particularly apparent from  around 550\,Myr ago --- the start of the Phanerozoic eon --- when hard-shelled animals first appeared and were preserved in the fossil record. Since then we observe a general increase in the diversity of life, but with significant variation superimposed \citep{sepkoski02, rohde05, alroy08sci}. The largest and most rapid decreases in biodiversity --- defined here are the number of genera extant at any one time --- are referred to as mass extinctions.

The cause of these variations in biodiversity in general, and mass extinctions in particular, has been the subject of intense study and speculation for over a century. Many mechanisms have been proposed for the observed variation, which we can place into four groups.

First, the variations are the result of inter-species interactions. Species
compete for limited resources, and as one species evolves to compete in this
struggle for survival, so other species will evolve too. This idea has been referred to as the ``Red Queen hypothesis'' (\cite{van-valen73, benton09}; in reference to the Red Queen's race in Lewis Carroll's {\em Through the Looking Glass}, where Alice must run just to keep still). Recent ecological studies indicate that the interaction between species can minimize competition and enhance biodiversity \citep{sugihara09}, although it is not obvious that these biotic factors are the main cause of large-scale patterns of biodiversity \citep{benton09, alroy08}.

Second, the environment changes with time, and species will evolve in response
to this. This ideas is sometimes called the ``Court Jester hypothesis''
\citep{barnosky01,benton09}.  Some of these (abiotic) geological changes are
relatively slow, such as plate tectonics, atmospheric composition, and global climate \citep{sigurdsson88, crowley88, wignall01, marti05, feulner09, wignall09}. Others may be more rapid. Large-scale volcanism, for example, would inject dust, sulfate aerosols and carbon dioxide into the atmosphere, resulting in a short-term global cooling, reduced photosynthesis, long periods of acid rain, and resulting ultimately in a long-term global warming (on a timescale of $10^5$ yr; \cite{marti05}).

Third, extraterrestrial mechanisms could be involved, either through a direct impact on life or by changing the terrestrial climate. Variations in Earth's orbit (mostly its eccentricity) over ten to one hundred thousand year time-scales are responsible for the ice ages \citep{hays76, muller00}. Extraterrestrial mechanisms on longer time scales could also play a role. These include solar variability \citep{shaviv03,lockwood05,lockwood07}, asteroid or comet impacts \citep{shoemaker83,alvarez80,glen94}, cosmic rays \citep{shaviv05, sloan08}, supernovae (SNe) and gamma-ray burst (GRBs; \cite{ellis95,melott08,domainko13}; for a review see \cite{bailer-jones09}). For example, cosmic rays might influence Earth's climate if they play a significant role in cloud formation (through the formation of cloud condensation nuclei; \cite{carslaw02,kirkby08}).
Secondary muons resulting from cosmic rays -- as well as high energy gamma rays
from SNe -- could kill organisms directly or damage their DNA \citep{thorsett95, scalo02, atri12}.

Finally, the apparent variation may be, in part, the result of uneven preservation and sampling bias \citep{raup72, alroy96, peters05}. Fossilization is relatively rare and some animals are more likely to be preserved than others.  Furthermore, the degree of preservation of marine spieces (more common in the fossil record) depends on the amount of continental outcrop available at any time, and this depends on the sea level \citep{hallam89, holland12}.  The number of species or genera living at any one time is not observed but must be reconstructed from the times at which species appear and disappear, which implies some kind of sampling or modeling. This can introduce a bias, although it may be diminished to some degree by various techniques \citep{alroy01, alroy10}.

These four types of cause of biodiversity variation are not mutually exclusive. They probably all acted at some point, and will also have interacted. For example, an asteroid impact could release so much carbon dioxide that long-term global warming has the biggest impact on biodiversity. Alternatively, a cool period would lower sea levels, leaving less continental shelf for the preservation of marine fossils, even though biodiversity itself may be unchanged.

Given the limited geological record, untangling the relevance of these different causes in most individual cases is difficult, if not impossible. More promising might be an attempt to identify the overall, long-term significance of these potential causes. The goal of this article is to do that for extraterrestrial phenomena. 

Many of the astronomical mechanisms mentioned above are ultimately caused by the presence of nearby stars. Stars turn SNe, the source of gamma rays, and their remnants are a major source of cosmic rays \citep{koyama95}. Stars perturb the Oort cloud, the main source of comets in the inner solar system \citep{rampino84,sanchez01}. Broadly speaking, when the Sun is in regions of higher stellar density, it is more exposed to extraterrestrial mechanisms of biodiversity change. In its orbit around the Galaxy (once every 200--250\,Myr or so), the Sun's environment changes. For example, it oscillates about the Galactic plane with a (quasi) period of 50--75\,Myr \citep{bahcall85}, and in doing so moves through regions of more intense star formation activity in the Galactic plane. This is particularly true if the Sun crosses spiral arms \citep{gies05, leitch98}, which it may do every 100-200\,Myr or so.

Such changes in the solar environment have been used as the basis for many
claims of a causal connection between the solar motion and mass extinctions
and/or climate change. Typically, authors have identified a periodicity in the
fossil record and then connected this to a plausible periodicity in the solar
motion \citep{alvarez84, raup84, davis84, muller88, shaviv03, rohde05,
  melott10, melott12}. These comparisons are fraught with problems, however,
some of which remained unmentioned by the authors. The first is the fact that
the solar motion and past environment are poorly constrained by the
astronomical data, so a wide range of plausible periods are permissible
\citep{overholt09, mishurov11}, yet the coincident one is naturally
chosen. Second comes the fact that the solar motion is not strictly periodic
even under the best assumptions. Third, many of these studies have not
performed a careful model comparison. Typically they identify the best--fitting period assuming the periodic model to be true, but fail to accept that a non-periodic model might explain the data even better\citep{kitchell84, stigler87}. In some cases a significance test is introduced to exclude a specific noise model, but this is often misinterpreted, and the resulting significance overestimated.  The reader is referred to \cite{bailer-jones09} for an in-depth review and references.

\subsection{Overview}\label{sec:overview}

Here we attempt a more systematic assessment of the possible role of the solar
orbit in modulating extraterrestrial extinction mechanisms. Our approach is
new in a number of respects, because we: (1) do a numerical reconstruction of
the solar orbit (rather than just assuming it to be periodic); (2) take into
account the observational uncertainties in that reconstruction; (3) use models
which predict the variation of probability of extinction with time (rather
than assuming that extinction events occur deterministically, for instance);
(4) do proper model comparison (rather than using {\it p}-values in an over-simplified significance test); (5) compare not only the orbital model with the fossil record but also numerous reference (non-orbital) models, such as periodic, quasi-periodic, trend, periodic with constant background, etc.

Our method is as follows. Adopting a model for the distribution of mass in the Galaxy, we reconstruct the solar orbit over the past 550\,Myr by integrating the Sun's trajectory back in time from the current phase space coordinates (position and velocity). This gives us a time series of how the stellar density in the vicinity of the Sun has varied over the Phanerozoic. We then assume that this density is approximately proportional to the terrestrial extinction probability (per unit time). That is, we adopt a non-specific kill mechanism linking the solar motion to terrestrial biodiveristy.  This is naturally a strong and rather simple assumption, but it should be emphasized that we are interested in the overall plausibility of extraterrestrial phenomena rather than trying to identify a specific cause of individual extinction events. The resulting time series is then compared to several different reconstructions of the biodiversity record.

A significant source of uncertainty in the reconstructed solar orbit is the current phase space coordinates (or ``initial'' conditions). We therefore sample over these to build up a set of (thousands of) possible solar orbits, and compare each of these with the data. The comparison is done by calculating the likelihood of the data for each orbit. Rather than finding the single most likely orbit, we calculate the average likelihood over all orbits. This is important, because it properly takes into account the uncertainties (whereas selecting the single most likely orbit would ignore them entirely). Indeed, these initial conditions can be considered as the six parameters of this orbital model (for a fixed Galactic mass distribution). We are, therefore, averaging the likelihood for this model over the prior plausibility of each of its parameters. This average --- or marginal --- likelihood is often called the ``evidence''.  This is just the standard, Bayesian approach to model assessment, which avoids the various flaws of hypothesis testing \citep{jeffreys61, winkler72, mackay03}, but unfortunately it has seen little use in this field of research.

The next step is to compare this evidence with that calculated for various
reference models (sometimes called ``noise'' or ``background'' models,
depending on the context). One such model is a purely sinusoidal model,
parameterized by an amplitude, period and phase. We generate a large number of
realizations of the model for different combinations of the parameters,
calculate the likelihood of the data for each, and average the results. This
averaging plays the crucial role of accommodating the complexity of the
model. A complex model with lots of parameters can often be made to fit an
arbitrary data set well. That is, it will give a high maximum likelihood. But
this does {\em not} make it a good model, precisely because we know that it
could have been made to fit any data set well! Such models are highly tuned,
so while the maximum likelihood fit may be very good, a small perturbation of
the parameters results in poor predictions. Unless supported by the data very
well, such models are less plausible. A simpler model, in contrast, may not
give such an optimal fit, but it is typically more robust to small
perturbations of the model parameters or the data, so gives good fit over a
wider portion of the parameter space. The model evidence embodies and
quantifies this trade off, which is why it --- rather than the maximum likelihood --- should be used to compare models.

We have selected four data sets for our study. The first two are compilations of the variation of extinction rate over time, from \cite{rohde05} and \cite{alroy08sci}. In the latter we use the extinction rate standardized to remove the sampling bias. Both report a magnitude as a function of time.
The second two data sets just record the time of mass extinction events. Here we take the times of the ``big 5'' mass extinctions and 18 mass extinctions identified by \cite{bambach06} based on Sepkoski's earlier work \citep{sepkoski86}.  Each mass extinction is represented as a (normalized) Gaussian on the time axis, the mean representing the best estimate of the date of the event and the standard deviation the uncertainty. We refer to these as ``discrete'' data sets, as they just list the discrete dates at which events occur (we do not use any magnitude information). The two rate data sets we therefore refer to as ``continuous'' (even though in practice the rates are also recorded at discrete time points).

The paper is arranged as follows. We first introduce the data sets. In Section~\ref{sec:model} we explain the modeling approach, how we calculate the likelihoods and evidence, and how we compare the models.  In Section~\ref{sec:solarorbit} we describe how we reconstructed the solar orbit, and also quantify the degree of periodicity typically present (as a strict periodicity has often been assumed in the past).  In Section~\ref{sec:result} we calculate and compare the evidences for the various models and data sets and test the sensitivity of the results to the model parameters and uncertainties in the data.  We conclude in Section~\ref{sec:conclusion}.

\section{Paleontological data}\label{sec:paleontological} 

We adopt four data sets: two discrete time series (sequence of time points
with age uncertainties) giving the dates of mass extinctions, and two
continuous time series giving the smoothed and normalized extinction rate as a
function of time.

\subsection{Discrete data sets}\label{sec:datadiscrete}

Sepkoski and others have identified five extinction events to be ``mass
extinctions'' \citep{sepkoski86}, often referred to as the ``big five''. Other
studies have identified different candidates for these, or have identified a
``big $N$'' for some other value of $N$. For example, Bambach et al.\ identify
three mass extinction events as being globally distinct \citep{bambach04}.
Here we adopt a set of 18 mass extinction events (or B18) selected by
\cite{bambach06} using an updated Sepkoski genus-level database. They are
consistently identifiable in different biodiversity data sets and when using
different tabulation methods. The second of our discrete data sets is the
``big five'' as identified from among the B18. Other choices of events are of
course possible and our results will, in general, depend on this choice
(although as we will see the results are rather consistent).

\begin{figure*}
  \centering
  \includegraphics[scale=0.7]{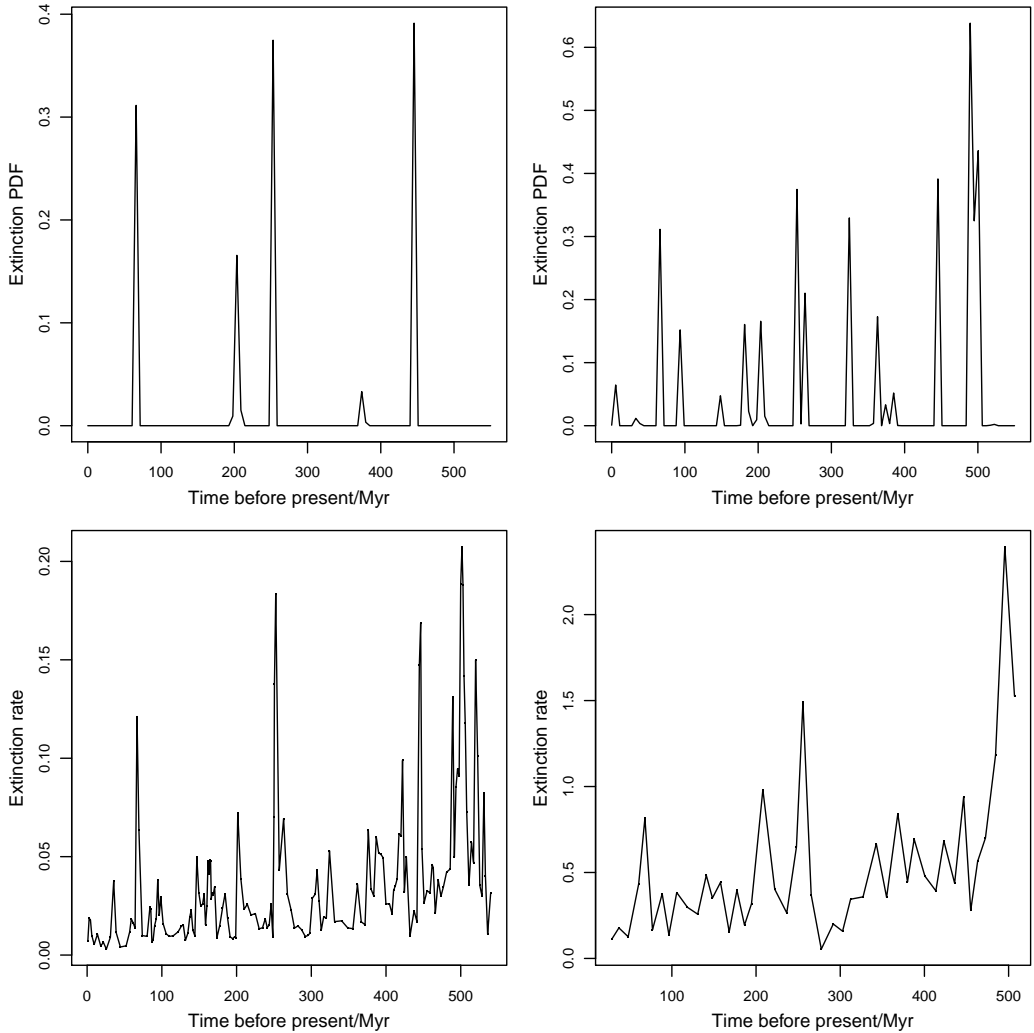}
  \caption{Four data sets used in this study. The top row shows the discrete data sets: the B5 (left) and B18 (right) mass extinction events. These can be interpreted as the extinction probability density function (PDF), which is proportional to the extinction fraction per unit time (i.e.\ a rate).  The bottom row shows the continuous data sets, which give the extinction rate: RM (left) and A08 (right).}
\label{fig:data} 
\end{figure*}

The times and durations of the events are listed in
Table~\ref{tab:ext-b5-b18}.  The time, $\tau$, is the mid-point between the
start age and the end age of the substage in which the extinction occurred, and the substage duration, $d$, is the difference between these. The geological record does not resolve the extinction event, so the extinction presumably  took place more rapidly than this substage duration. In that case $\tau$ is our best estimate of the true (but unknown) time, $t$, at which the extinction occurred, and $d$ is a measure of our uncertainty in this estimate. Uncertainty is represented by probability, so we interpret an ``event'' as the probability distribution $P(\tau | t, d)$. This is the probability that we would measure the event time as $\tau$, given $t$ and $d$.  We could represent this as a rectangular (``top-hat'') distribution of mean $t$ and width $d$, but this assigns exactly zero probability outside the substage duration, which implies certainty of the start and end ages.  Even though their (relative) ages have uncertainties which are less the
  event duration, we nonetheless accommodate some uncertainty in these start
  and end ages by considering each event to be a Gaussian distribution with
mean $t$, and standard deviation $\sigma$ equal to the standard deviation of
the rectangular distribution, which is $\sigma = d/\sqrt{12}$.  This
  Gaussian distribution is broader than the corresponding rectangular distribution.  Note that the intensity of the mass extinction is not taken into account. A Gaussian is normalized, so its peak value is determined by its standard deviation (Figure~\ref{fig:data}).\footnote{One might think that a more natural interpretation of an event is $P(t | \tau, d)$, the probability that the true event occurs at $t$ given the measurements. But here we are considering the measurement model (or noise model), that is, given some true time of the event, what possible times might we measure, the discrepancy arising on account of the finite precision of our measurement process.}

\begin{table}
\caption{The B18 mass extinction events, with the B5 shown in bold. BP = before present.}
\centering
\begin{tabular}{l c}
\hline
\hline
Time ($\tau$) / Myr BP & Substage duration ($d$) / Myr\\
\hline
3.565&3.53\\
35.550&3.30\\
{\bf 66.775}&{\bf 2.55}\\
94.515&2.03\\
146.825&2.65\\
182.800&7.00\\
{\bf 203.750}&{\bf 8.30}\\ 
{\bf 252.400}&{\bf 2.80}\\
262.950&5.10\\
324.325&4.15\\
361.600&4.80\\
{\bf 376.300}&{\bf 3.60}\\
386.925&3.25\\
{\bf 445.465}&{\bf 3.53}\\
489.350&2.10\\
495.725&2.15\\
499.950&2.10\\
519.875&2.75\\
\hline 
\end{tabular}
\label{tab:ext-b5-b18}
\end{table}

\subsection{Continuous data sets}\label{sec:continuous} 

The discrete data sets are naturally biased in that they only select periods of high extinction rate. It may well be that extraterrestrial phenomena are only relevant in causing (or contributing to) mass extinctions, but a priori it is natural to ask how the overall extinction rate varies.  The extinction rate, $E(t)$, is the fraction of genera which go extinct in a stratigraphic substage divided by its duration. This is directly proportional to the variation of extinction probability per unit time.  For one of our extinction rate data sets we use the linearized and interpolated data set constructed by \cite{rohde05} as reported in \cite{bambach06}. We denote this RM. The other data set is the ``three-timer'' extinction rate from the Paleobiology database\footnote{paleodb.org} \citep{alroy08sci}. The data are binned into 48 intervals averaging 11\,Myr in duration. The counts are derived from 281\,491 occurrences of 18\,541 genera within 42\,627 fossil collections. We use the data set processed using their subsampling method in order to reduce the sampling bias, and denote this A08.  Both data sets are reported as lists of extinction rates at specific times, $\{r_j, \tau_j\}$. These two continuous data sets are plotted in the lower row of Figure~\ref{fig:data}.

\section{Data modeling method}\label{sec:model}

\subsection{Overview of Bayesian model comparison}\label{sec:overview} 

The goal of this work is to compare how well various models predict the paleontological data sets.  We do this by calculating, for a given data set, the Bayesian evidence for each model. If the models are equally probable a priori, then the one with the highest evidence is the best predictor of the data. This does not exclude the possibility that there exists a better model which we have not yet tested. But it at least allows us to conclude that the lower evidence models are neither appropriate nor sufficient explanations of the phenomenon. (Indeed, we never assume a model is ``true'', just better than the alternatives.)

The modelling approach is described in full by \cite{bailer-jones11, bailer-jones11-err}, so will only be outlined here. Let $D$ denote the paleontolgical time series, and $M$ the model. Examples of $M$ are the orbital model or the periodic model.  Each model has a set of parameters, $\theta$. The {\em likelihood} of the model, $P(D | \theta, M)$, is the probability of obtaining $D$ from model $M$ with its parameters set to some specific values of $\theta$.
Normally we do not know the exact values of these parameters, and the data --- being noisy and imperfectly fit by the model --- do not determine them exactly either. We therefore average the likelihood over all possible values of $\theta$, weighting each by how plausible that value of $\theta$ is. This weighted average is the evidence.
This weighting is given by the prior probability distribution, $P(\theta | M)$.
In the case of the orbital model, where $\theta$ is the current phase space coordinates of the Sun,
$P(\theta | M)$ is determined by the observational uncertainties in these measurements.
Mathematically the evidence is
\begin{equation} 
  \label{eqn:evidence} 
  P(D|M)=\int_\theta P(D|\theta,M) P(\theta|M) d\theta \ \ .  
\end{equation}
It gives the probability of getting the data from the model, regardless of the specific values of the parameters, i.e.\ it measures how well the model explains the data. The absolute value of the evidence is not of interest, so we generally deal with the ratio of two evidences for two models, known as the {\em Bayes factor}.
The evidence is a far better measure of the suitability of a model than is the
{\it p}-value \citep{jeffreys61, kass95, winkler72, mackay03, bailer-jones09}. 

It is worth stressing again that, by averaging over the parameters, the evidence is not sensitive to model complexity per se. This is in contrast to the likelihood at the best fitting parameters (the maximum likelihood): a more complex (flexible) model will always fit the data better, and so will always deliver a higher maximum likelihood. The evidence reports the average likelihood, so it will only increase if the extra complexity gives a net benefit over the plausible parameter space. The model complexity does not then need to be considered separately in some ad hoc way.

\subsection{The likelihood}\label{sec:principle} 

\subsubsection{Discrete data sets}\label{sec:likediscrete} 

As explained in Section~\ref{sec:datadiscrete}, each event in a discrete data set can be considered 
as a Gaussian distribution. For event $j$ this gives the probability that the mass extinction occurs at time $\tau_j$, given that the true time is $t_j$ and the uncertainty in our measurement is $\sigma_j$. That is,
\begin{equation}
  \label{eqn:measurement}
  P(\tau_j | \sigma_j, t_j) = \frac{1}{\sqrt{2\pi}\sigma_j} \, e^{-(\tau_j - t_j)^2/2\sigma_j^2} \ \ .
\end{equation}
In order to compare the measurement of this event with the predictions of a model we calculate the likelihood. This is given by an integral over the unknown true time
\begin{eqnarray}
  \label{eqn:likelihood_event}
  P(\tau_j | \sigma_j, \theta, M)&=&\int_{t_j} P(\tau_j | \sigma_j, t_j,
\theta, M) P(t_j | \sigma_j, \theta, M) \rm{d} t_j \nonumber \\
                            &=&\int_{t_j} P(\tau_j | \sigma_j, t_j) P(t_j | \theta, M) \rm{d}t_j \ .
\end{eqnarray}
The first term in the integral -- Eqn.~\ref{eqn:measurement} -- 
is sometimes called the measurement model. The second term is the prediction
of the time series model, i.e.\ the probability (per unit time) that a mass
extinction occurs at time $t_j$. Our time series models are, therefore,
stochastic in the sense that they do not attempt to predict when mass
extinctions occurred, but rather how the probability of occurrence of a mass
extinction varies over time. Note that the likelihood is just measuring the
degree of overlap between the data and the model predictions, averaged over
all time.

Eqn.~\ref{eqn:likelihood_event} gives the likelihood for a single event. Assuming all events are measured independently\footnote{ More precisely, the events are assumed independent given the model and its parameters. This is probably a reasonable assumption given that the events are distributed quite sparsely over the Phanerozoic, and that the separations between them are generally much longer than their substage durations.}, then the likelihood for all the data is just the product of the event likelihoods
\begin{equation}
  \label{eqn:likelihood_disc}
  P(D|\theta, M) = \prod_j P(\tau_j| \sigma_j,\theta, M)
\end{equation}
where $D=\{\tau_j\}$. For the sake of the likelihood and evidence calculation we do not consider the $\{\sigma_j\}$ as data, although they are of course measured. That is because $D$ is defined as just those quantities which are predicted by the measurement model.

\subsubsection{Continuous data sets}\label{sec:continuous} 

Mathematically the likelihood for the continuous data is very similar as in the discrete case, but the interpretation is  different. 

Consider the measurement model, $P(\tau | \sigma, t)$, in Eqn.~\ref{eqn:measurement} (we consider just one event so drop the subscript $j$).  We have interpreted this as the probability of a discrete extinction event being measured at $\tau$, but we could equivalently interpret it as the probability density (i.e.\ probability per unit time) of extinction at time $\tau$. Now, rather than characterizing the probability density as a Gaussian with mean $t$ and standard deviation $\sigma$, we could consider an arbitrary function, characterized by a series of top-hat functions, $\{p_i\}$ (a histogram), each top-hat characterized by a center $t_i$, height $r_i$, and width $\delta_i$.  We can then replace $P(\tau | \sigma, t)$ with $\sum_i p_i(\tau | t_i, \delta_i)$ where
\begin{eqnarray}
   p_i(\tau | t_i, \delta_i) =  
\left\{
     \begin{array}{ll}
          r_i & \mbox{\rm when}~~ t_i - \delta_i/2 < \tau < t_i + \delta_i/2\\
          0   & \mbox{\rm otherwise \ .}
     \end{array}
\right.
\end{eqnarray}
We now see that
\begin{equation}
\lim_{\delta_i \rightarrow 0} \sum_i p_i(\tau | t_i, \delta_i) = E(t)
\end{equation}
i.e.\ we get a continuous function of the variation of the extinction probability with time $t$ (as $\tau$ and $t$ become equivalent in this limit). The likelihood is then
\begin{equation}
  \label{eqn:likelihood_cont}
  P(D|\theta, M) = \int_{t} E(t) P(t | \theta, M) \rm{d}t \ .
\end{equation}
In practice we characterize $E(t)$ using the extinction rate, $r_j$, tabulated at each time $\tau_j$, which is equivalent to assuming that extinction rate is constant over the substage (or that we have zero uncertainties on the measured times). 

We can actually apply this interpretation to the discrete data sets too. In both cases, the data provide the variation of extinction probability (per unit time) as a function of time, or something proportional to that. The proportionality constant is irrelevant, because we keep the data fixed when comparing different models using the evidence.

\subsection{Time series models}\label{sec:tsmodels} 

\begin{table*}
\centering
\caption{The mathematical form of the time series models and their corresponding parameters. Time $t$ increases into the past and $P_u(t|\theta, M)$ is the unnormalized extinction probability density predicted by the model.}
\begin{tabular}{l c r}
\hline
\hline
model name&$P_u(t|\theta,M)$&parameters\\
\hline
Uniform&1&none\\
RNB/RB&$\sum_{n=1}^{N}\mathcal{N}(t; \mu_n,\sigma)$+$B$&$\sigma$, $N$, $B$\\
PNB/PB&$1/2\{\cos[2\pi(t/T+\beta)]+1\}$+$B$&$T$, $\beta$, $B$\\
QPM&$1/2\{\cos[2\pi t/T + A_Q \cos(2 \pi t/T_Q) +\beta]+1\}$&$T$, $\beta$, $A_Q$, $T_Q$\\
SP&$[1+e^{(t-t_0)/\lambda}]^{-1}$&$\lambda$, $t_0$\\
SSP&PNB+SP&$T$, $\beta$, $\lambda$, $t_0$\\
OM(P)/SOM(P)&$n(\overrightarrow{r_\odot}(t),\overrightarrow{v_\odot}(t))$&$\overrightarrow{r_\odot}(t=0),~\overrightarrow{v_\odot}(t=0)$\\
\hline
\end{tabular}
\label{tab:models}
\end{table*}

\begin{table*}
\centering
\caption{Range of parameters adopted in the model prior parameter distributions.
Except for OM(P)/SOM(P), a uniform prior for all parameters for all models is adopted which is constant inside the range shown, and zero outside. The prior PDF of parameters in the OM(P)/SOM(P) model is Gaussian and specified by the uncertainties in the initial conditions.}
\begin{tabular}{l c}
\hline
\hline
model name & range of prior \\
\hline
Uniform& None\\
RNB/RB & $\sigma=10$~Myr, $N\in\{5,18\}$, $B\in\{0,\frac{1}{\sqrt{2\pi}\sigma}\}$\\
PNB/PB & $10<T<100$, $0<\beta<2\pi$, $B\in[0,1]$\\
QPM &$10<T<100$, $0<\beta<2\pi$, $0<A_Q<0.5$, $200<T_Q<500$\\
SP &$-100<\lambda<100$, $100<t_0<500$\\ 
SSP &$10<T<100$, $0<\beta<2\pi$, $-100<\lambda<100$, $100<t_0<500$\\ 
OM(P)/SOM(P) & initial conditions (see Table~\ref{tab:initialconditions})\\
\hline
\end{tabular}
\label{tab:prior}
\end{table*}

The time series models appear in the equation for the likelihood in the form $P(t | \theta, M)$, i.e.\ the extinction probability (per unit time) as a function of time predicted by model $M$ at parameters $\theta$. 
It is important to realize that this probability density function (PDF) over $t$
is normalized, i.e.\ integrates over all time to unity. This is key to model comparison, because a model which assigns a lot of probability to extinctions at some particular time must necessarily assign lower probability elsewhere. This follows because we are not trying to model the absolute value of the extinction rate, but just its relative variations.

In addition to specifying the functional form of the models we must also specify the prior probability distribution of the model parameters, $P(\theta | M)$. This describes our prior knowledge of the relative
probability of different parameter settings. For example, given the time scale
in the data, we are not interested in models with time scales less than a few
million years or more than a few hundred million years. It is often difficult to be precise about priors, and the evidence and therefore Bayes factors often depend on the choice. 
The choice of prior must therefore be considered part of the model (e.g.\
``periodic model with permissible periods between 10 and 100\,Myr'' is
distinct from ``periodic model with permissible periods between 50 and
60\,Myr''). We  investigate the sensitivity of the results to
changes in the prior in Section~\ref{sec:sensitivity}.
Except for the orbital model, we adopt a uniform prior over all model parameters over the range specified in Table~\ref{tab:prior}.

\begin{figure}
\centering
\includegraphics[scale=0.7]{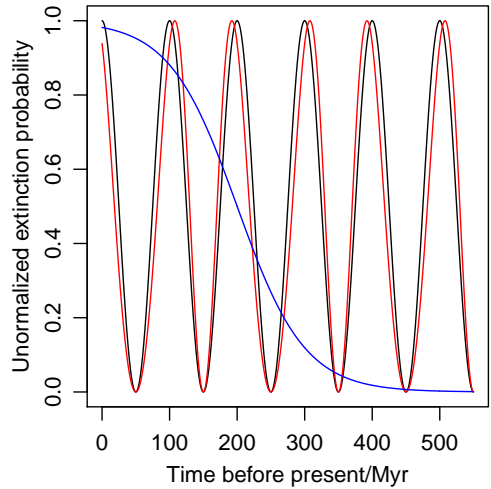}
\caption{Time series models. The black line shows the PNB model with
  $T=$\,100\,Myr and $\beta=0$. The red line shows the QPM model with $T=$\,100\,Myr, $\beta=0$, $A_Q=0.5$ and $T_Q=$\,200\,Myr; The blue line shows the SP model with $\lambda=$\,50\,Myr and $t_0=$\,200\,Myr.}
\label{fig:models}
\end{figure}

Table~\ref{tab:models} summarizes the functional form of the models, which are now briefly described. 
Figure~\ref{fig:models} plots examples of some of these models.
The range of the data is taken to be 0--550\,Myr BP.
\begin{description}
\item[Uniform] Constant extinction PDF over the range of the data. This has no parameters.

\item[RB/RNB] Random model in which a set of $N$ times is drawn at random from a uniform distribution extending over the range of the data. A Gaussian with standard deviation $\sigma=$\,10\,Myr is assigned to each of these, and then a constant $B$ added before normalizing. This is the RB model.  The RNB (``random no background'') model is just the special case of $B=0$, which produces a model which is similar to our discrete data. 
In practice we fix $N$ and $B$ and calculate the evidence by averaging over a large number of realizations of the model. Specifically, when modelling the B5 and B18 data sets we fix $B=0$, and $N=5,18$ respectively. 

\item[PB/PNB] Periodic model of period $T$ and phase $\beta$ (model PNB). There is no amplitude parameter because the model is normalized over the time span of the data. Adding a background $B$ to this simulates a periodic variation on top of a constant extinction probability (model PB).

\item[QPM] A quasi-periodic model in which the phase is a sinusoid with amplitude $A_Q$, period $T_Q$ and phase $\theta$ (it becomes the same as the PNB model if $A_Q=0$). 

\item[SP] A monotonically increasing or decreasing nonlinear trend in the extinction PDF using a sigmoidal function characterized by the steepness of the slope, $\lambda$ and the center of the slope, $t_0$. In the limit that $\lambda$ becomes zero the model becomes a step function at $t_0$, and in the limit of very large $\lambda$ becomes the uniform model.

\item[SSP] Combination of SP and PNB.

\item[OM(P)/SOM(P)] The orbital/semi-orbital model with/without spiral arms, defined in Section~\ref{sec:OBM}.
\end{description}

\subsection{Numerical calculation of the evidence}\label{sec:numcalc}

The integral in Eqn.~\ref{eqn:evidence} is a multidimensional integral over the parameter space, and cannot be calculated analytically. As in \cite{bailer-jones11}, we estimate it using a Monte Carlo method, by drawing model parameters at random from the prior distribution and calculating the likelihood at each. If the set of $N$ parameter draws is denoted $\{\theta\}$, then Eqn.~\ref{eqn:evidence} can be approximated as the average likelihood
\begin{equation}
  \label{eqn:evidencenum}
  P(D | M) \, \simeq \, \frac{1}{N}  \sum_{\theta} P(D | \theta, M) \ \ .
\end{equation}
In the following simulations we adopt $N=$\,10\,000 unless noted otherwise.

\section{Model of the solar orbit}\label{sec:solarorbit} 

We now reconstruct the orbit of the Sun around the Galaxy over the past 550\,Myr. This is done by integrating the Sun's path back in time through a fixed gravitational potential, described in Section~\ref{sec:potential}. (The dynamics are reversible because only gravity acts; energy is not dissipated.) It has often been assumed that the solar orbit is periodic with respect to crossings of the Galactic plane and/or spiral arms; we investigate this numerically in Section~\ref{sec:periodicity}.  The stellar mass distribution corresponding to the potential gives the local stellar density which the Sun experiences in its orbit. In Section~\ref{sec:OBM} we use this to derive the variation in the expected extinction rate.

\subsection{The Galactic potential}\label{sec:potential}

We adopt an analytic Galaxy potential, $\Phi_G(R,z)$ (described in cylindrical coordinates), comprising three components
\begin{equation}
  \label{eqn:potential}
  \Phi_G=\Phi_b+\Phi_h+\Phi_d\ . 
\end{equation}
The first two components are spherically symmetric distributions which represent the bulge and halo using Plummer's model \citep{plummer11}
\begin{equation}
  \label{eqn:pot-bh}
  \Phi_{b,h}=-\frac{G~M_{b,h}}{\sqrt{R^2+z^2+b_{b,h}^2}}
\end{equation}
and the third is an axisymmetric disk according to \cite{miyamoto75}
\begin{equation}
  \label{eqn:pot-d}
  \Phi_d=-\frac{G~M_d}{\sqrt{R^2+(a_d+\sqrt{z^2+b_d^2})^2}} \ .
\end{equation}
In Eqn.~\ref{eqn:pot-bh} and \ref{eqn:pot-d} $R$ is the galactocentric radius
and $z$ is the distance from the Galactic midplane. $M_b$, $M_h$ and $M_d$ are
the mass of bulge, halo and disk, respectively. $a_d$ and $b_d$ are the scale
length and height (respectively) of the disk, and $b_d$ and $b_h$ are the scale
lengths of the bulge and halo respectively. We adopt the numerical values for
these as given in Table~\ref{tab:modelpar} (see \citep{sanchez01}). 
In addition to the these three components, we introduce into some of the
models a time-dependent potential due to the Galactic spiral arms, denoted
$\Phi_s(R,\phi,z,t)$. It is defined below in Section~\ref{sec:arm}.

\begin{center}
\begin{table}
\caption{The parameters of Galactic potential model (from \cite{sanchez01})}
\centering
\begin{tabular}{ll}
\hline
\hline
component & parameter value \\\hline
Bulge    & $M_b=1.3955 \times 10^{10}~M_\odot$ \\
                   & $b_b =0.35$\,kpc \\
Halo     & $M_h=6.9766\times 10^{11}~M_\odot$\\
                   & $b_h=24.0$\,kpc\\
Disk     & $M_d=7.9080\times 10^{10}~M_\odot$\\
                   & $a_d=3.55$\,kpc\\
                   & $b_d=0.25$\,kpc\\
\hline
\end{tabular}
\label{tab:modelpar}
\end{table}
\end{center}

There is, of course, significant uncertainty not only in the value of the parameters of the potential, but also in the functional form of our model. It is no doubt a simplification of the true potential of the Galaxy, so specific numerical values of quantities such as orbital periods and amplitudes inferred should not be interpreted too literally.
Below we investigate the sensitivity of our model comparison to changes in the potential.

\subsection{Orbit calculation}

To calculate the motion of a body through the potential from given initial conditions, we solve Newton's equations of motion, which in cylindrical coordinates are
\begin{eqnarray}
\ddot{R} -R\dot\phi^2 &=& -\frac{\partial\Phi}{\partial R} \nonumber \\
R^2 \ddot{\phi}+2R\dot{R}\dot{\phi} &=& -\frac{\partial\Phi}{\partial\phi} \nonumber \\
\ddot{z} &=& -\frac{\partial \Phi}{\partial z} 
\end{eqnarray}
We solve these equations by numerical integration using the {\tt lsoda} method
implemented in the {\it R} package {\tt deSolve}, with a time step of 0.1\,Myr. 

The initial conditions are the current phase space coordinates (three spatial
and three velocity coordinates) of the Sun. These are derived from
observations with a finite accuracy, so our initial conditions are Gaussian
distributions, with mean equal to the estimated coordinate and standard
deviation equal to its uncertainty (Table~\ref{tab:initialconditions}). In
order to calculate an orbit we draw the initial conditions at random
from these prior distributions, and a large number of draws gives us a
sampling of orbits which will be used later (e.g.\ in the evidence calculations).

We derive our initial conditions from a number of sources:
The distance to the Galactic centre comes from astrometric and spectroscopic
observations of the stars near the black hole of the Galaxy
\citep{eisenhauer03}. The Sun's displacement from the galactic plane is
calculated from the photometric observations of classical Cepheids by
\cite{majaess09}. The Sun's velocity is calculated from Hipparcos data by \cite{dehnen98b}.

\begin{table*}
\caption{The current phase space coordinates of the Sun, represented as
  Gaussian distributions, and used as the initial conditions in our orbital
  model}
\centering
\begin{tabular}{l*5{c}r}
\hline
\hline
     &$R$/kpc&$V_R$/kpc~Myr$^{-1}$&$\phi$/rad&$\dot\phi$/rad~Myr$^{-1}$&z/kpc&$V_z$
/kpc~Myr$^{-1}$\\
\hline
mean &8.0  &-0.01              & 0        & 0.0275
&0.026&0.00717\\
standard deviation &0.5& 0.00036&0         &0.003
&0.003&0.00038\\
\hline
\end{tabular}
\label{tab:initialconditions}
\end{table*}

\subsection{Spiral arms}\label{sec:arm}

\begin{table*}
\caption{The parameters of the geometric and potential model for the spiral arms. The parameters
of the potential apply to both arms.}
\centering
\begin{tabular}{lcccc}
\hline
\hline
\multicolumn{5}{c}{geometric parameters}\\\hline
arm&$\alpha$&$R_{min}$/kpc&$\phi_{min}$/rad&extent/kpc\\\hline
1        & 4.25 & 3.48 & 0.26 & 6.0\\
$1^\prime$&4.25  & 3.48 & 3.40 & 6.0\\\hline
\hline
\multicolumn{5}{c}{potential parameters}\\\hline
N&$\rho_0/M_\odot$\,kpc$^{-3}$&$r_0$/kpc&$R_s$/kpc&H/kpc\\\hline
2&$2.5\times 10^7$&8&7&0.18\\\hline
\end{tabular}
\label{tab:spiralarmpar}
\end{table*}

The model for the spiral arms is described by their geometry and their gravitational potential. 
However, for the arm crossing periodicity test in the next section we ignore mass of the spiral arms when calculating the solar orbit and only consider their location. 
Likewise, in one of the class of variants of the orbital model, OM and SOM (defined later), we ignore the arms
entirely (for both the orbit and stellar density calculations).
This is done so that we can see the additional effect of the arms, the form and mass of which are poorly determined by current observations.

\begin{figure*}
\centering
\includegraphics[scale=0.7]{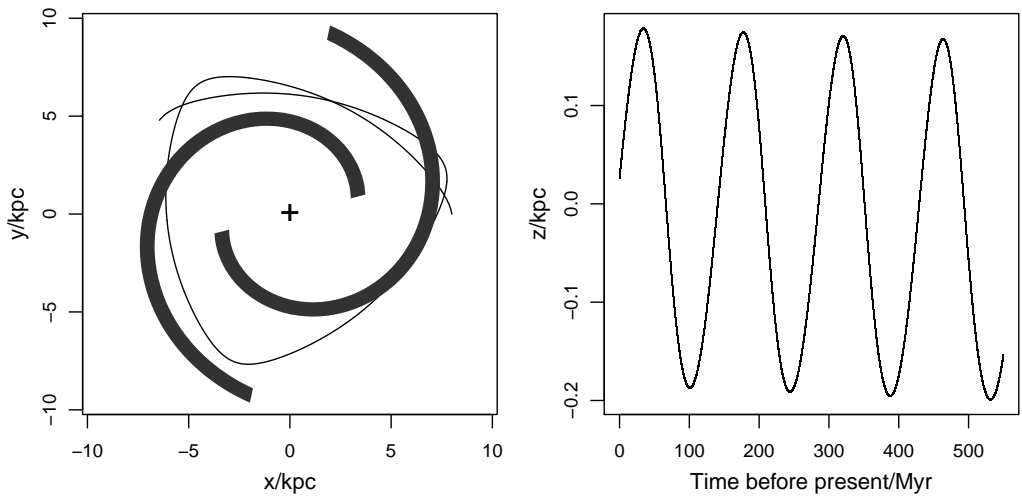}
\caption{The Solar orbit in the Galactic plane (left, thin lines) and perpendicular to the
  plane (right). The orbit in the left panel is in a reference frame rotating with the spiral arms (shown in thick lines).} 
\label{fig:spiralzorbit} 
\end{figure*}

The geometric model comprises two logarithmic spiral arms, the positions of which in circular coordinates, $(R, \phi)$, are given by
\begin{equation}
  \label{eqn:logspiral}
  \phi_s(R)=\alpha \log(R/R_{min})+\phi_{min},
\end{equation}
where $\alpha$ is a winding constant, $R_{min}$ is the inner radius and
$\phi_{min}$ is the azimuth at that inner radius. The radius of the spiral arm
ranges from $R_{min}$ to $R_{max}$. Of the various arm models offered by
\cite{wainscoat92}, we selected the main two spiral arms, 1 and $1^\prime$,
with $\phi_{min}$ given by \cite{vanhollebeke09} and other parameters given by
\cite{wainscoat92} (see Table~\ref{tab:spiralarmpar}).  Their location in the plane of the Galaxy is shown in the left panel of Figure~\ref{fig:spiralzorbit}. The arms rotate rigidly with constant angular velocity (pattern speed) of $\Omega_p=20$\,km\,s$^{-1}$\,kpc$^{-1}$ \citep{martos04, drimmel00}.

The gravitational potential of the arms is described by the (first term of
the) analytic potential of Eqn.~8 of \cite{cox02}. This is
\begin{eqnarray}
  \label{eqn:Phis}
  \Phi_s=& -\frac{4\pi GH}{K_1 D_1} \rho_0 e^{-\frac{R-r_0}{R_s}} \times \nonumber\\
     &\cos(N[\phi-\phi_s(R,t)])\left[{\rm sech}\left(\frac{K_1 z}{\beta_1}\right)\right]^{\beta_1} \ ,
\end{eqnarray}
where 
\begin{eqnarray*}
K_1 &=& \frac{N}{R \sin(\alpha)},\\
\beta_1 &=& K_1 H(1+0.4K_1H),\\
D_1&=& \frac{1+K_1H+0.3(K_1H)^2}{1+0.3K_1H},\\
\phi_s(R,t)&=&\phi_s(R)+\Omega_p t \ . \\
\end{eqnarray*}
The amplitude of the spiral density distribution is
\begin{equation}
\rho_A(R, z) = \rho_0 e^{-\frac{R-r_0}{R_s}} {\rm sech}^2(z/H) \ ,
\end{equation}
where $R_s$ is the radial length of the drop-off in density amplitude of the
arms, $\rho_0$ is the midplane arm density at fiducial radius $r_0$, and $H$
is the scale height of the spiral density. This modulated by a sinusoidal pattern in
$\phi$, to give the overall density which corresponds approximately\footnote{The density corresponding strictly to the potential in Eqn.~\ref{eqn:Phis} is rather messy, and is not necessary for this work. The approximate density in Eqn.~\ref{eqn:rhos} is accurate enough for our purposes for
 small $H/r$ and $z/H$ not too large \citep{cox02}. }
to the above potential
\begin{equation}
  \label{eqn:rhos}
  \rho_s(R,\phi, z, t) = \rho_A(R,z)\cos[N(\phi-\phi_s(R,t))],
\end{equation} 
where N is the number of arms and $\phi_s(R)$ is given in Eqn.~\ref{eqn:logspiral}. The values of the relevant parameters are given in Table~\ref{tab:spiralarmpar}. 

\subsection{Periodicity test}\label{sec:periodicity} 

In previous studies of the impact of astronomical phenomena on the terrestrial biosphere, it has frequently been assumed that the solar motion shows strict periodicities in its motion perpendicular to the Galactic plane, and sometimes also with respect to spiral arm crossings. We investigate this here using our numerical model.

For each orbit $k$, we calculate the intervals between successive crossings, $\{\Delta t^i\}_k$, (separately for midplane and spiral arm crossings), where $i$ indexes the crossing. We then calculate the sample mean and sample standard deviation of these intervals for each orbit
\begin{eqnarray}
\overline{\Delta t_k} &=& \frac{1}{N_k-1}\sum_{i=1}^{N_k-1} \Delta t_k^i\\
\sigma_k &=& \sqrt{\frac{1}{N_k-2} \sum_{i=1}^{N_k-1} (\Delta t_k^i-\overline{\Delta t_k})^2} \ ,
\end{eqnarray}
where $N_k$ is the number of crossings in the $k$th orbit. To assess the periodicity of the crossing
intervals, we define the degree of {\em aperiodicity} as
\begin{equation}
a_k=\sigma_k/\overline{\Delta t_k} \ .
\end{equation}
An orbit with $a=0$ is strictly periodic.

We investigated the variation of the aperiodicity of the solar orbit with the six parameters (initial conditions).  This parameter space is too large to report on extensively here, but we find that the aperiodicity is most sensitive to $R(t=0)$ and $\dot\phi(t=0)$.
In the following we vary these initial conditions individually, by drawing $10^4$ samples from the corresponding initial condition distribution. (Larger sample sizes did not alter the results significantly)
We simulate the solar orbits using the arm-free potential, $\Phi_G$.

\subsubsection{Midplane crossings}\label{sec:midplane} 

Some earlier studies claimed that Galactic midplane crossings trigger increases in terrestrial extinction due to an enhanced gamma ray or cosmic ray flux or due to larger perturbation of the Oort Cloud. These are directly related to 
the increased stellar density and increased occurrence of star-forming
regions. The larger tidal forces are postulated to enhance the disruption
of the Oort cloud \citep{rampino00, matese95}, and the higher density of massive stars --- and thus high energy radiation as well as increased SN rate --- raises the average flux the Earth is exposed to.
The periodicity of the Sun's vertical motion -- not least its period, phase and the assumed stability of this period -- is central to these claims.  We examine these using our model.

\begin{figure*}
\centering
\includegraphics[scale=0.7]{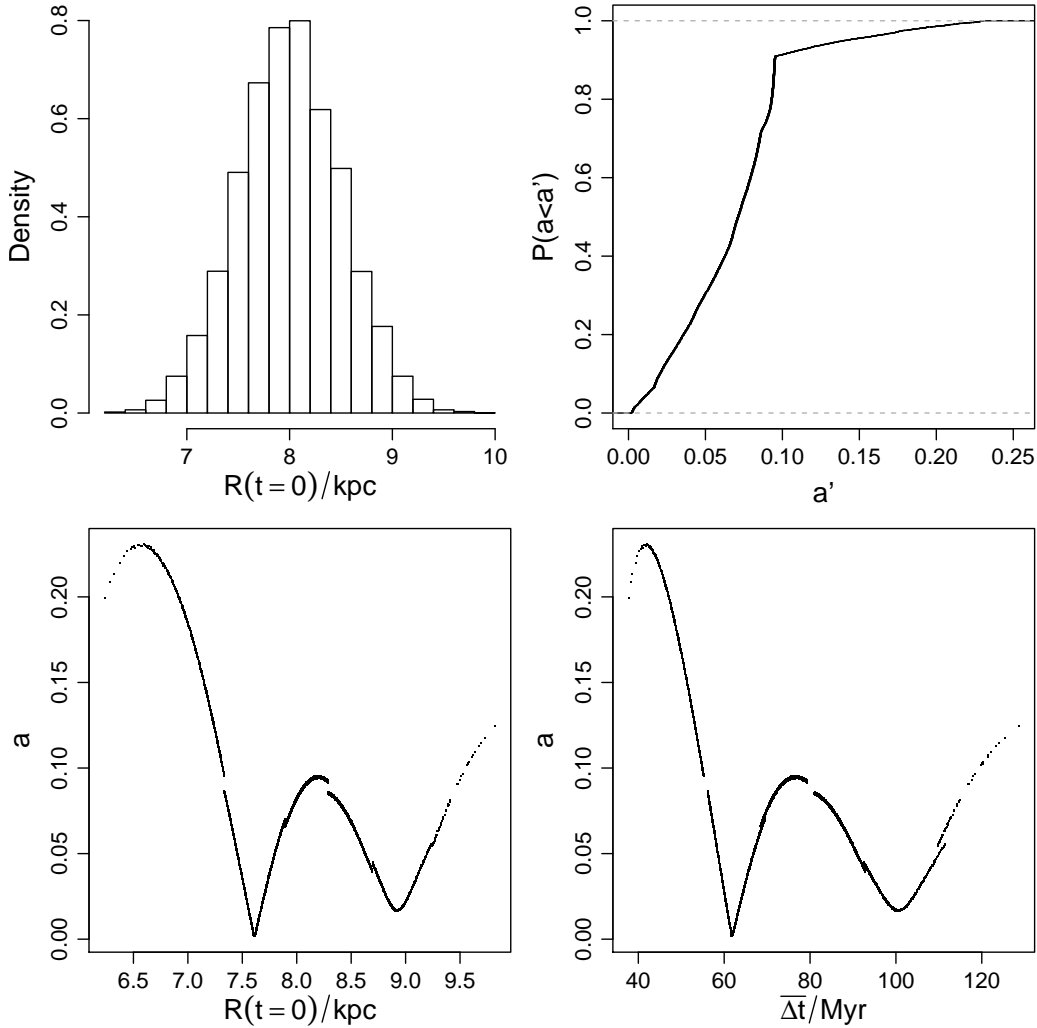}
\caption{Periodicity test of midplane crossings varying just the current galactocentric radius of the Sun, $R(t=0)$.
 Top left: distribution of this initial condition in the simulations (it has a Gaussian distribution with parameters given in Table~\ref{tab:initialconditions}). Top right: cumulative probability of the aperiodicity parameter for the resulting orbits. Bottom left: the variation of aperiodicity with $R(t=0)$. Bottom right: the variation of aperiodicity with the average crossing interval, $\overline{\Delta t_k}$.}
\label{fig:sen-varyR-plane} 
\end{figure*}

The results of varying just the initial galactocentric radius of the Sun, $R(t=0)$, are shown in Figure~\ref{fig:sen-varyR-plane}. We see in the top-right panel that about 90\% orbits have an aperiodicity less than 0.1. In the lower two panels we see how $a$ varies with the value of the initial condition and with the average crossing interval.

The aperiodicity is 0.002 (nearly strict periodicity) at $\overline{\Delta t}=61.8$\,Myr.  
This corresponds to a 1:1 resonance between the vertical motion and the radial
motion. Its value is close to a period in the biodiversity data of $62\pm 3$\,Myr claimed by \cite{rohde05}. Little should be read into this coincidence, however, as there is no good (i.e.\ independent) reason to select the specific initial condition that leads to this period over any other. 
Moreover, changing the parameters of the Galactic potential --- which is not very well known --- changes this period. 
(For example, if we increase the mass of the Galactic halo the values of
$\overline{\Delta t}$ are decreased.) The other minimum in the aperiodicity in
the bottom panels is 0.02 at $\overline{\Delta t}=100.2$\,Myr. This
corresponds to an approximately circular orbit in the midplane. If we set $V_R(t=0)=0$, $V_z(t=0)=0$ and $z=0$, this solar orbit would be strictly circular.

The cumulative curve (top-right panel of Figure~\ref{fig:sen-varyR-plane}) makes a sharp turn at $a'=0.1$. This is because of a sudden decrease in the number of orbits with large aperiodicities.  Similarly, the discontinuities in the lower panels are caused by changes in the (small) number of discrete plane crossings which occur for different aperiodicity ranges.

If we now vary the initial condition $\dot\phi(t=0)$ instead, the periodicity test gives very similar results: we find a nearly strict periodicity at $\overline{\Delta t}=60$\,Myr and another minimum in the aperiodicity at about 100\,Myr. That means the nearly strict periodicity is mainly determined by a combination of $R(t=0)$ and $\dot\phi(t=0)$.

In summary, we see that the majority of the simulated orbits (90\%) are quite close to periodic ($a\leq0.1$) in their motion vertical to the midplane, although strict periodicity essentially never occurs.
 
\subsubsection{Spiral arm crossings}\label{sec:spiral} 

Regarding spiral arms as regions of increased star formation activity and
stellar density, the mechanisms of mass extinction considered for midplane
crossing could likewise be applied to spiral arm crossings, and have been by
some authors \citep{leitch98, gies05}. However, such studies have over simplified the solar motion by failing to take into account the considerable uncertainties in the current phase space coordinates of the Sun and thus in its plausible orbits.  Some studies have even claimed a connection between spiral arm crossings and the terrestrial biosphere {\em after} having fit the solar motion to the geological data, but such reasoning is clearly circular.

We examine here the periodicity of spiral arm crossings (although we note that
some studies in the literature claiming a spiral arm-extinction link just
consider the crossing times and do not claim a periodic crossing).  The
crossing intervals are longer than with the midplane, so we only include in
our analyses models in which there are at least three arm crossings. We assume
that the arms have indefinite vertical extent, so that a crossing on the {\it x--y} plane is always a true encounter. In reality the Sun might pass over or under the arms, thus reducing the overall relevance of spiral arm crossings to terrestrial extinction.

\begin{figure*}
\centering
\includegraphics[scale=0.7]{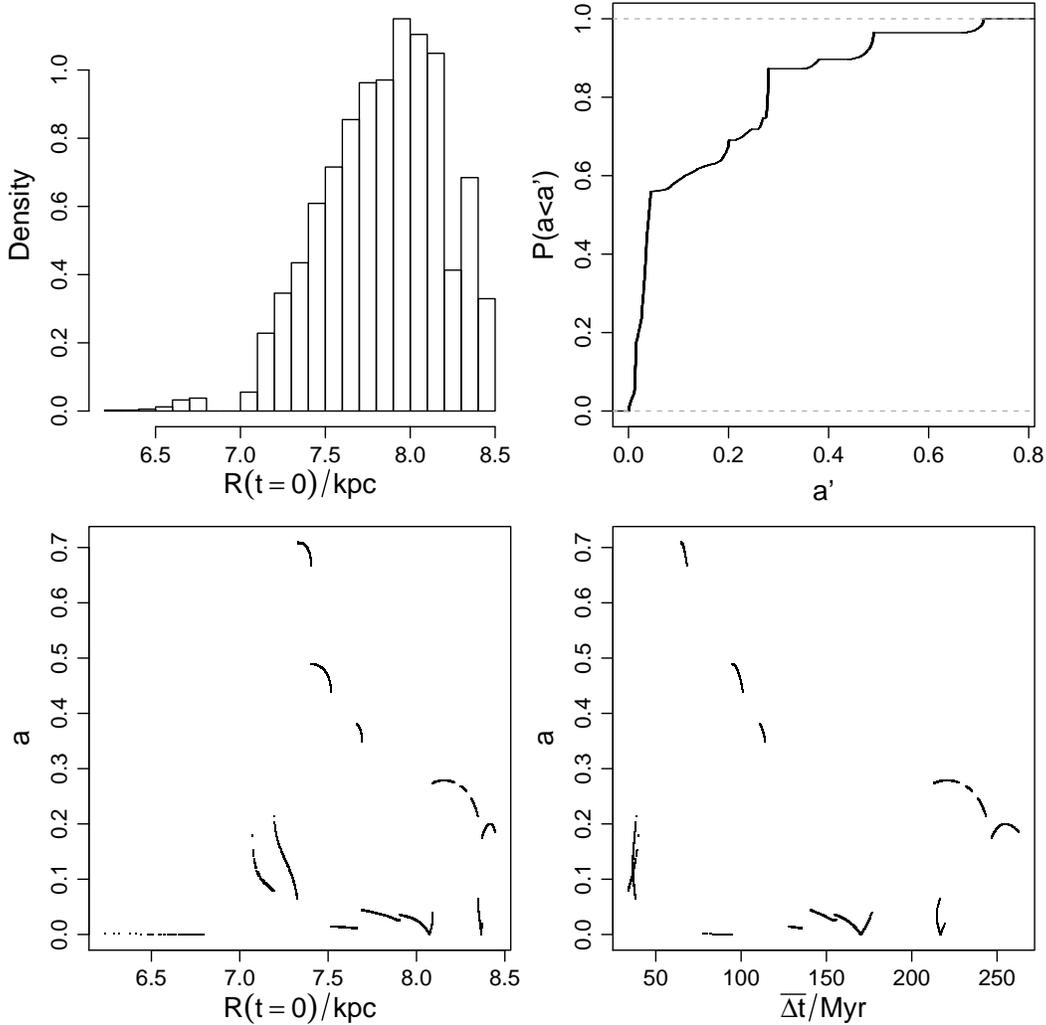}
\caption{Same as Figure~\ref{fig:sen-varyR-plane}, but now for the spiral arm crossings.}
\label{fig:sen-varyR-arm} 
\end{figure*}

Figure~\ref{fig:sen-varyR-arm} shows the result of this analysis for the 7\,407 orbits (out of the original sample of 10\,000) that exhibit at least three arm crossings.  The cumulative probability (top-right panel) shows that about 40\% of the orbits have an aperiodicity larger than 0.2. In other words, it is not very likely that the solar orbit and spiral arms are so tuned to give periodic crossings.  The lower two graphs show how $a$ varies with $R(t=0)$ and $\overline{\Delta t_k}$.  The numerous gaps in these plots are a consequence of the fact that not all orbits for certain ranges of $R(t=0)$ had at least three arm crossings, and so were removed from the analysis. We see, therefore, that the crossing interval is very sensitive to $R(t=0)$.

Note that we have neglected the mass of the arms in the orbital calculations. When we include it the values of aperiodicity increase and there is an even less clear dependence of $a$ on 
$R(t=0)$ or $\overline{\Delta t_k}$.

In summary, we find it unlikely that spiral arm crossings are even close to periodic. If the pattern speed of the spiral arms has not been constant in the past 550\,Myr, or if the pattern itself has not been stable, then this conclusion is strengthened further.

\subsection{Orbital model}\label{sec:OBM}

\subsubsection{Derivation of the extinction rate from the stellar density variation}

As outlined in Section~\ref{sec:introduction}, various astronomical mechanisms for biological extinction have been identified, including comet impacts (from Oort Cloud perturbation), gamma rays (from SNe or GRBs), and cosmic rays (from SN remnants; \cite{ellis95,sanchez01,gies05}).  The intensity of all of these depends on the local stellar density. If we consider a general mechanism involving flux from nearby stars, then the flux from a single star is proportional to $f/d^2$, where $f$ is the relevant surface flux and $d$ the distance. The sum of this over the whole relevant volume of space around the Sun is proportional to the total intensity and thus the extinction probability (per unit time). 

Let us assume that the extinction rate, $E$, is linearly proportional to the flux, and that the number density of relevant stars is proportional to the total stellar number density (stars per unit volume), $n$.  Because the density of spiral arms is much less than the density of the other components, we consider at first only the time-independent density arising from halo, disk and bulge.  The density is calculated from the corresponding potential (defined in Section~\ref{sec:potential}) using Poisson's equation.

In an axisymmetric cylindrical coordinate system, the extinction rate at the Sun is then
\begin{eqnarray}
  \label{eqn:stellar_density1}
  E(R_\odot, z_\odot) &=& C \int \!\! \int \frac{n(R,z)}{d^2} \,R\,dR\,dz\nonumber\\
                           &=&C \int \!\! \int \frac{n(R,z)}{(R-R_\odot)^2+(z-z_\odot)^2}\,R\,dR\,dz,\nonumber\\
\end{eqnarray}
where $R$ and $z$ are the galactocentric radius and height above the midplane,
respectively, for some star, $R_\odot$ and $z_\odot$ are the corresponding
(time-varying) coordinates of the Sun, and $C$ is a constant. Note that the
stellar number density, $n$, is proportional to the corresponding stellar density, $\rho$. Defining the distance from a star to the Sun as
$r\equiv R-R_\odot$ and $Z\equiv z-z_\odot$, the extinction rate is  
\begin{eqnarray}
  \label{eqn:stellar_density2}
  E(R_\odot,z_\odot) &=& C \int \!\! \int\frac{n(R_\odot+r,z_\odot+Z)}{r^2+Z^2}(R_\odot+r) drdZ \ .\nonumber\\
\end{eqnarray}
The flux from a star falls off as $1/d^2$, but we can truncate this integral at some upper distance because
at some point the flux is too weak to influence the terrestrial biosphere. We
take $d_{\rm th}=50$\,pc as an upper limit.\footnote{In the case of SNe, \cite{ellis95} conclude that only those which come within 10\,pc of the Sun would have a significant impact on terrestrial life.
GRBs up to 1\,kpc or even more could still have an effect on the
Earth, but we ignore these because the GRB rate (at low redshifts) is comparatively low (e.g.\ \cite{domainko13}).}
This is much smaller than the scale
length of the disk and comparable to the scale height of the disk (see Table~\ref{tab:modelpar}), so we can approximate $n(R_\odot+r,z_\odot+Z)(R_\odot+r)$ by
$n(R_\odot,z_\odot+Z)R_\odot$. The integral then becomes
\begin{equation}
  \label{eqn:extinctionPDF1}
  E(R_\odot,z_\odot)\simeq CR_\odot\int_{-d_{\rm th}}^{d_{\rm th}}\int_{-d_{\rm th}}^{d_{\rm th}}\frac{n(R_\odot,z_\odot+Z)}{r^2+Z^2} drdZ \ .
\end{equation}
Integrating over $r$ gives
\begin{equation}
  \label{eqn:stellar_density3}
  E(R_\odot,z_\odot)\simeq 2CR_\odot\int_{-d_{\rm th}}^{d_{\rm th}}n(R_\odot,Z+z_\odot)\frac{\arctan(d_{\rm th}/Z)}{Z}\,dZ \ .
\end{equation}
The geometric factor $\frac{\arctan(d_{\rm th}/Z)}{Z}$ drops close to zero at about $Z=$\,25\,pc, and does so much more rapidly than the stellar density term, which follows the vertical profile of the disk (which has a much larger scale height of 250\,pc).
Thus to a reasonable degree of approximation we can set $n(R_\odot, Z+z_\odot)
\simeq n(R_\odot, z_\odot)$ in this integral. The integral is then just over
the geometric factor, which gives some constant (dependent on $d_{\rm th}$,
but of no further interest).
Thus we are left with 
\begin{equation}
  \label{eqn:stellar_density4}
  E(R_\odot,z_\odot)\simeq C'R_\odot n(R_\odot,z_\odot),
\end{equation}
for some constant $C'$. For the solar motion, the relative
variation of $R_\odot$ is less than that of $n(R_\odot,z_\odot)$, so we have
\begin{equation}
  \label{eqn:stellar_density5}
  E(R_\odot,z_\odot) \propto n(R_\odot, z_\odot) \ .
\end{equation}
In other words, the extinction rate is just proportional to the stellar
density at the location of the Sun. 
The approximations in Eqns.~\ref{eqn:extinctionPDF1}--\ref{eqn:stellar_density5} still hold  when we include the low density spiral arms defined in Section~\ref{sec:arm}, in which case we must also introduce the explicit dependence on azimuth and time
\begin{equation}
E(R_\odot, \phi_\odot, z_\odot, t) \propto n(R_\odot,\phi_\odot,z_\odot,t) \ .
\end{equation}

In the above model we assumed that the extinction rate is proportional to
$d^{-2}$, i.e.\ the influence falls off like a flux on the surface of a
sphere. We could generalize this dependence to be $d^{-k/2}$ for $k\geq0$ in
order to reflect other mechanisms, e.g.\ tidal effects. 

\begin{figure*}
\centering
\psfrag{Rs}[c][c]{\scriptsize{$R_\odot$/kpc}}
\psfrag{Zs}[c][c]{\scriptsize{$z_\odot$/kpc}}
\includegraphics[scale=0.7]{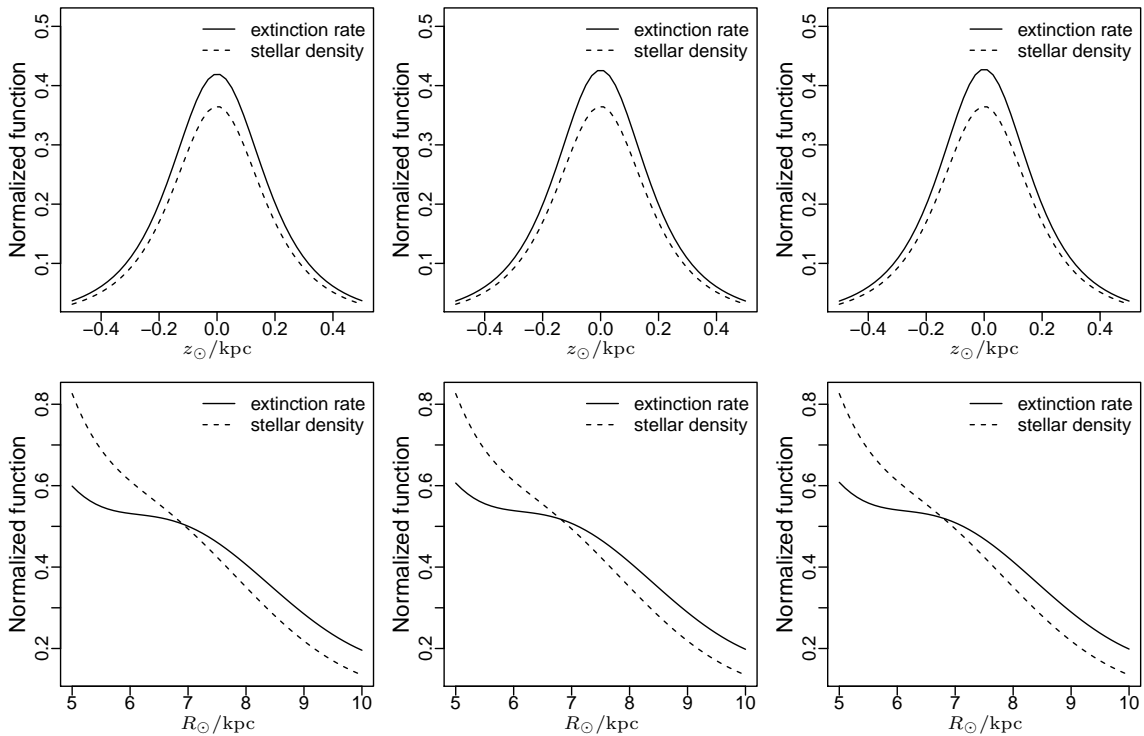}
\caption{Comparison of the extinction rate calculated numerically with the stellar density at the position of the Sun.
The top row shows the variation as a function of $z_\odot$ with $R_\odot$ fixed to 8\,kpc. The bottom row show the variation as a function of $R_\odot$ with $z_\odot$ fixed to 26\,pc. The columns from left to right are for $k=0,2,4$ in the model for the dependence of extinction rate with distance.}
\label{fig:ext-pos10}
\end{figure*}

In order to test the validity of the above approximations, we compare in Figure~\ref{fig:ext-pos10}
the extinction rate as given by Eqn.~\ref{eqn:stellar_density1} (by numerical
integration) with the stellar number density $n(R_\odot, z_\odot)$. We plot over
ranges of $R_\odot$ from 5\,kpc to 10\,kpc and $z_\odot$ from $-0.5$\,kpc to
0.5\,kpc, in accordance with the ranges covered by the simulated solar orbits.
We normalize the extinction rate (and the stellar density)
by setting its integral over $R_\odot$ and $z_\odot$ to be unity.
In the upper row of Figure~\ref{fig:ext-pos10}, the difference between the stellar density and the extinction rate reaches a maximum in the midplane ($z_\odot=0$); this is on account of the relatively large density gradient at $z_\odot=0$. The maximum difference is only about 10\% of the peak value of stellar density for all values of $k$. 
In the lower row, the largest difference is at the lower limit of $R_\odot$.
Note that the value of $k$ has very little impact.

In practice, most of the simulated orbits spend most of their time in the region $7<R_\odot/{\rm kpc}<9$ and $-0.3<z_\odot/{\rm kpc}<0.3$, where the differences between local stellar density and extinction rate variation are even smaller. Thus to within a few percent, the stellar density at the Sun is a good predictor of the extinction rate. The time variation of this density is the time series model forms the basis for what we refer to as the ``orbital models'', the forms of which we now define.

\subsubsection{Definition of OM(P) and SOM(P)}

The orbital model ``OM'' is the orbital model which does not include the spiral arm at all, neither in the gravitational potential (for calculation of the orbits) nor in the stellar density (for the extinction rate calculation). The orbital model OMP does include the spiral arm in both senses. Thus both OM and OMP are internally self-consistent. 

Once normalized, $E(t)$ is just the quantity $P(t | \theta, M)$ in Section~\ref{sec:tsmodels} (and it is normalized to give unit integral over the span of the data). The parameters of OM and OMP are the initial conditions of the orbit, and the corresponding priors are the Gaussian distributions summarized in Table~\ref{tab:initialconditions}. Thus one orbit calculated from one draw of the initial conditions allows us to calculate one likelihood for these models (for given data set). Repeating this and averaging the resulting likelihoods gives the evidence for that orbital model (see Section~\ref{sec:numcalc}).

For both of these models we consider four variations, labeled 1--4, according to which initial conditions we vary (and therefore sample over to build up the set of orbits).

\begin{figure*}
\centering
\includegraphics[scale=0.7]{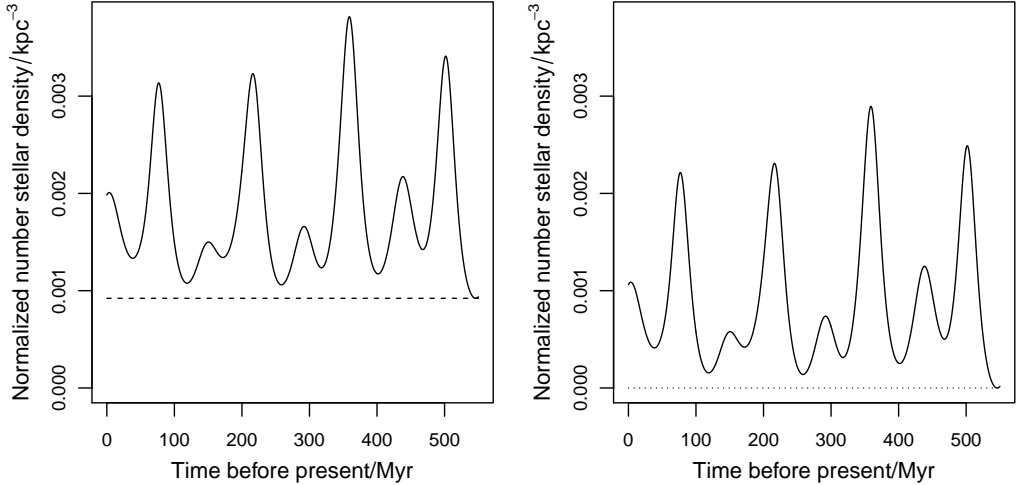}
\caption{Orbital model (OM) and semi-orbital model (SOM). The left
  panel shows the variation of the local stellar density -- and thus the
  extinction probability per unit time -- in one particular orbit calculated
  from OM. The horizontal dashed line is a threshold, $h$, for truncating the stellar density to a minimum level, which gives rise to the extinction probability
  per unit time plotted in the right panel.}
\label{fig:som}
\end{figure*}

In addition to these models, we define the``semi-orbital model'', SOM. This is derived from the OM simply by subtracting from the predicted extinction rate a constant value, $h$, and setting all resulting negative values to zero. Here we simply set $h$ to be the minimum value of the extinction rate (see Figure~\ref{fig:som}). This is intended to model the situation in which the flux causing the extinction must rise above some threshold before it has an effect. (We might consider this as an adaption of life to the extraterrestrial flux background.) In analogy to OM, SOM excludes the spiral arm. SOMP is SOM with the spiral arm potential and density included. Once again we will consider four varieties according to which initial conditions are varied.

\section{Results}\label{sec:result} 

\subsection{Evidences}\label{sec:evidences} 

\begin{table*}
\caption{Bayes factors and maximum likelihood ratios of the various time series
  models (rows) relative to the Uniform model for the various data sets
  (columns). OM(P)1--4 refer to the OM(P) model in which different initial
  conditions are varied: $R(t=0)$, $\phi\dot(t=0)$, $\{R(t=0),~\dot\phi(t=0)\}$, $\{R(t=0),~V_R(t=0),~\dot\phi(t=0),~V_z(t=0)\}$, respectively, and likewise for the SOM(P)1--4 models. The other initial conditions are kept fixed. The RB and RNB models are intrinsically discrete, so are not applied to the two continuous data sets.}
\centering
\begin{tabular}{l|llll|llll}
\hline
\hline
     &\multicolumn{4}{c|}{Bayes factor (BF)}&\multicolumn{4}{c}{Maximum
       likelihood ratio (MLR)}\\
\cline{2-9}
Model&B5   &B18   &RM  &A08 &B5 &B18 &RM  &A08\\
\hline
PNB &0.97 &0.62   &0.98&0.87&22 &255 &1.2&1.1\\
PB  &1.0  &0.80   &0.98&0.87&3.5&45  &1.1&0.99\\
QPM &0.99 &0.85   &0.98&0.87&6.8&35  &1.1&0.98\\
RNB &0.041&0.00050&--&--&1153&16 &--&--\\
RB  &0.85 &0.40   &--&--&9.0&8.4 &--&--\\
SP  &0.28 &0.019  &1.02&0.88&2.7&0.36&2.1&1.3\\
SSP &0.73 &0.18   &0.99&0.87&8.6&81  &1.3&1.2\\
OM1 &1.4  &0.74   &0.99&0.88&4.6&2.4 &1.1&1.0\\
OM2 &1.4  &0.72   &0.99&0.89&4.9&2.2 &1.2&1.0\\
OM3 &1.2  &0.63   &0.99&0.88&4.9&2.6 &1.3&1.1\\
OM4 &1.2  &0.65   &0.99&0.88&5.0&3.0 &1.2&1.1\\
OMP1&0.18 &0.014  &0.93&0.88&6.3&0.48&5.9&4.4\\
OMP4&0.14 &0.022  &0.93&0.83&20 &6.1 &6.0&5.0\\
SOM1&1.3  &0.051  &1.0 &0.90&11 &0.34&1.2&1.2\\
SOM2&0.85 &0.037  &1.0 &0.91&5.9&0.33&1.3&1.2\\
SOM3&0.99 &0.032  &1.0 &0.89&24 &0.67&1.4&1.2\\
SOM4&1.0  &0.032  &1.0 &0.89&28 &0.70&1.3&1.3\\
SOMP1&0.11&0.00013&0.94&0.88&3.5&0.0067&5.9&4.4\\
SOMP4&0.10&0.0012 &0.94&0.83&20 &1.8 &6.0 &5.1\\
\hline
\end{tabular}
\label{tab:BF-MLR}
\end{table*}

We now calculate the Bayesian evidence (Eqn.~\ref{eqn:evidence}) for the
various models for each data set. This is done by sampling from prior
probability distributions of the model parameters ($P(\theta | M$),
Table~\ref{tab:prior}), calculating the likelihood
(Eqn.~\ref{eqn:likelihood_disc} for discrete time series,
Eqn.~\ref{eqn:likelihood_cont} for continuous time series) and then
averaging these for that model and data set Eqn.~\ref{eqn:evidencenum}. 

To calculate the evidences for RNB and PNB models for the B5 and B18 data sets, we adopt a Monte Carlo sample size of $10^6$. In all other cases we use a sample size of $10^4$. Larger sample sizes did not alter the estimated evidence significantly.\footnote{ For all the data sets, the standard error of the Monte
    Carlo estimates of the evidence is $<1$\% for OM models, $<3$\%
    for SOM and other models, and $<25$\% for RNB and (S)OMP models.}  This sample size is given according to the sensitivity test of the evidence to the sample size for all models and all data sets. This test shows that the evidence estimated from $10^4$ draws in the prior distribution is close to the real evidence when there is a background either
in the model or in the data set. 

As the absolute value of the evidence is not of interest, we report the ratio of evidence, the Bayes factor.  Here we report Bayes factors with respect to the Uniform model. We regard a model as being significantly better than another when its evidence exceeds that of the other by a factor of 10 \citep{jeffreys61, kass95}. Note that it is only meaningful to compare evidences --- and therefore Bayes factors --- for a fixed data set.

The results are shown in Table~\ref{tab:BF-MLR}. For the reference models we
evaluate the evidence by sampling over all their model parameters, but in the
case of OM and SOM we sample over just some of the parameters (initial
conditions), keeping the others fixed, in order to investigate the impact of
the different parameters.  As shown in Section~\ref{sec:periodicity}, the
periodicity of solar orbit is most sensitive to the initial conditions
$R(t=0)$ and $\dot\phi(t=0)$. We therefore calculate the evidence for the OM
(and SOM) models with four different sets of initial conditions being varied:
$R(t=0)$ only; $\phi\dot(t=0)$ only; $\{R(t=0)$ and $\dot\phi(t=0)\}$;
$\{R(t=0)$, $V_R(t=0)$, $\dot\phi(t=0)$, and $V_z(t=0)\}$.  In all cases we
fix $\phi(t=0)$ and $z(t=0)$, the former because it has no impact on the solar
motion in this axisymmetric potential, and the latter because the uncertainty
in the current $z$ position of the Sun has a limited impact on the subsequent
orbit. To assess the effect of the spiral arm perturbation on the Bayes factors, we have selected four perturbed orbital models, OMP1, OMP4, SOMP1 and SOMP4, to compare with corresponding unperturbed orbital models.

For the B5 data set, the Bayes factors of all time series models relative to the Uniform model are less than 10. 
Thus none of these models are a significantly better explanation of the
data. One model, RNB, has a Bayes factor less than 0.1, indicating that we can
discount this one as being an unlikely explanation. Given that the Uniform
model is the simplest model of the set, the principle of parsimony suggests
that we should be satisfied with it as explanation. This does not deny the possibility that some other model shows significantly higher evidence. After all, we can only ever make claims about models which we explicitly test.

\begin{figure}
\centering
\includegraphics[scale=0.7]{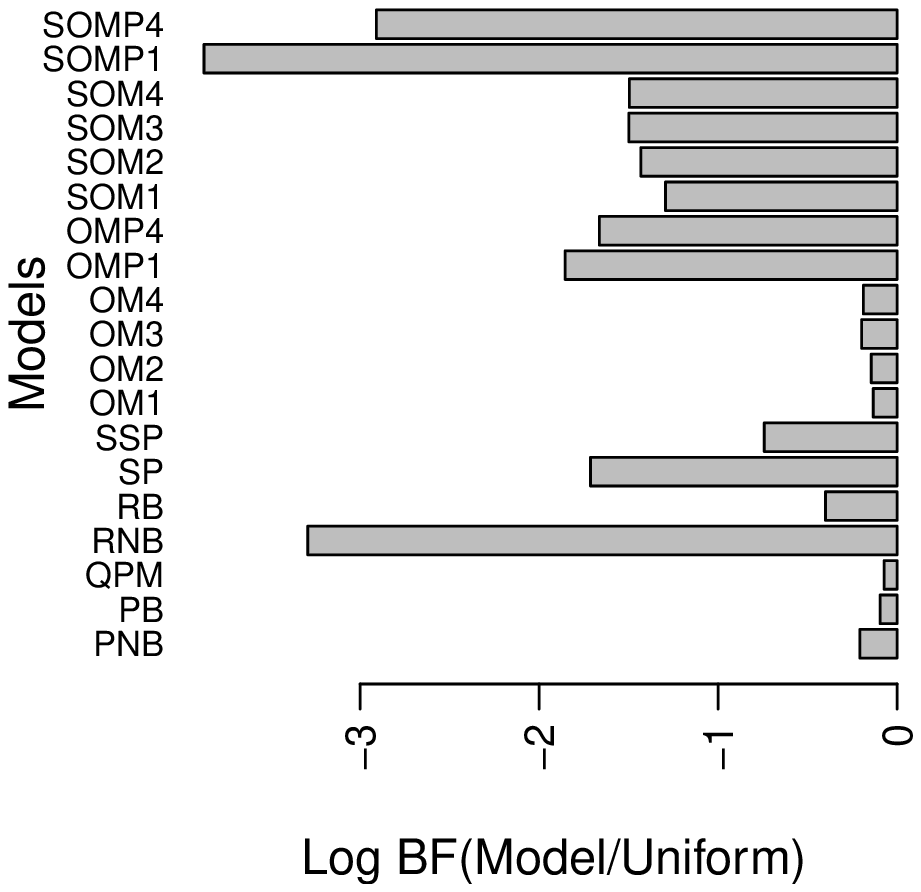}
\includegraphics[scale=0.7]{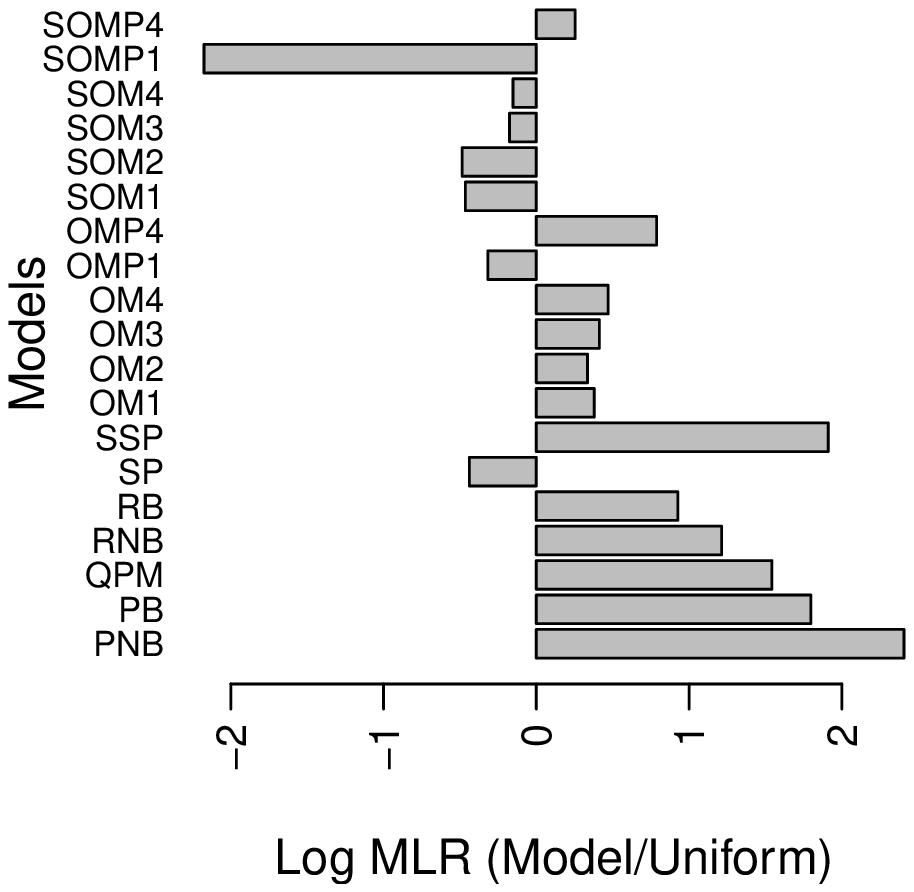}
\caption{Results for the B18 data sets. The upper panel shows the log (base 10)
  Bayes factor of the various models relative to the Uniform model. The lower panel shows the log (base 10) of the maximum likelihood ratio of various models relative to the Uniform model.
} 
\label{fig:BF-MLR}
\end{figure}

The B18 data set includes more extinction events than the B5 data set, and not
surprisingly it discriminates more between the
models (the Bayes factors show a larger spread). 
(These results are also shown graphically in the upper panel of Figure~\ref{fig:BF-MLR}.)
The OM models are favored somewhat more than the other models -- e.g.\ the
Bayes factor
of OM3 to SP is $0.63/0.019=33$ -- although again no model is favored
significantly more than the Uniform model. In contrast, several models are
significantly disfavored (RNB, SP, OMP1, OMP4, SOM1, SOM2, SOMP1 and SOMP4). 
In particular, the perturbed orbital models, including OMP1, OMP4, SOMP1, and
SOMP4, are less favored by the data than their corresponding unperturbed orbital
models. All the other perturbed orbital models (not listed in Table
\ref{tab:BF-MLR}) also have lower Bayes factors than the unperturbed orbital models. 

For the two continuous time series, RM and A08, the difference between the
evidences of all of the models is not significant.\footnote{ As the RNB and RB models are obviously conceptually inappropriate models for continuous data sets, we do not apply them to the A08 or RM data sets, so these values are missing from Table~\ref{tab:BF-MLR}.}
Our broad conclusion is that no model significantly outperforms the Uniform model on any of the data sets. On the contrary, a few can be ``rejected'' on the ground of a significantly lower evidence. Recall that all of these models are predicting the extinction probability (per unit time). In terms of the discrete data sets, the Uniform model just means that the mass extinction events occur at random in time. We find this to be no less probable than a periodic or quasi-periodic variation of the probability, or a monotonic trend in the probability, etc.  In terms of the continuous data sets, we obviously do not believe that the Uniform model is a good explanation of the clearly apparent variations in the extinction rate (see Figure~\ref{fig:data}). But the analysis does tell us that this is no worse an explanation than the more complex models of the variation considered, such as periodic, orbital-model based etc. Clearly there must be yet other models which could explain the data even better.  This may explain why previous authors have found an apparent periodicity in the data: the periodic model can explain the data to some degree, but actually no better than simpler models. 

\subsection{Likelihood distribution}\label{sec:likelihood}

We have seen that the evidence hardly discriminates between any of the models on the continuous data sets, and only between some of them on the discrete data sets. (This is by no means inevitable. In other problems the evidence can vary enormously between models.)  This means that, on average over their parameter space, the models differ little in their predictions.  It is nonetheless interesting to see how the likelihood varies over the parameter space.  (We would do this in particular to find the best fitting parameters, although these are only meaningful if the overall model has been identified as the best explanation of the data.)  We focus here mainly on the PNB and OM models for the B18 data. We again normalize the likelihood for a model by dividing it by the likelihood of the Uniform model to form the likelihood ratio.  As the latter model has no parameters, it is likelihood constant and equal to its evidence.
The maximum value of the likelihood ratio we denote as MLR.

\begin{figure}
\centering
\includegraphics[scale=0.7]{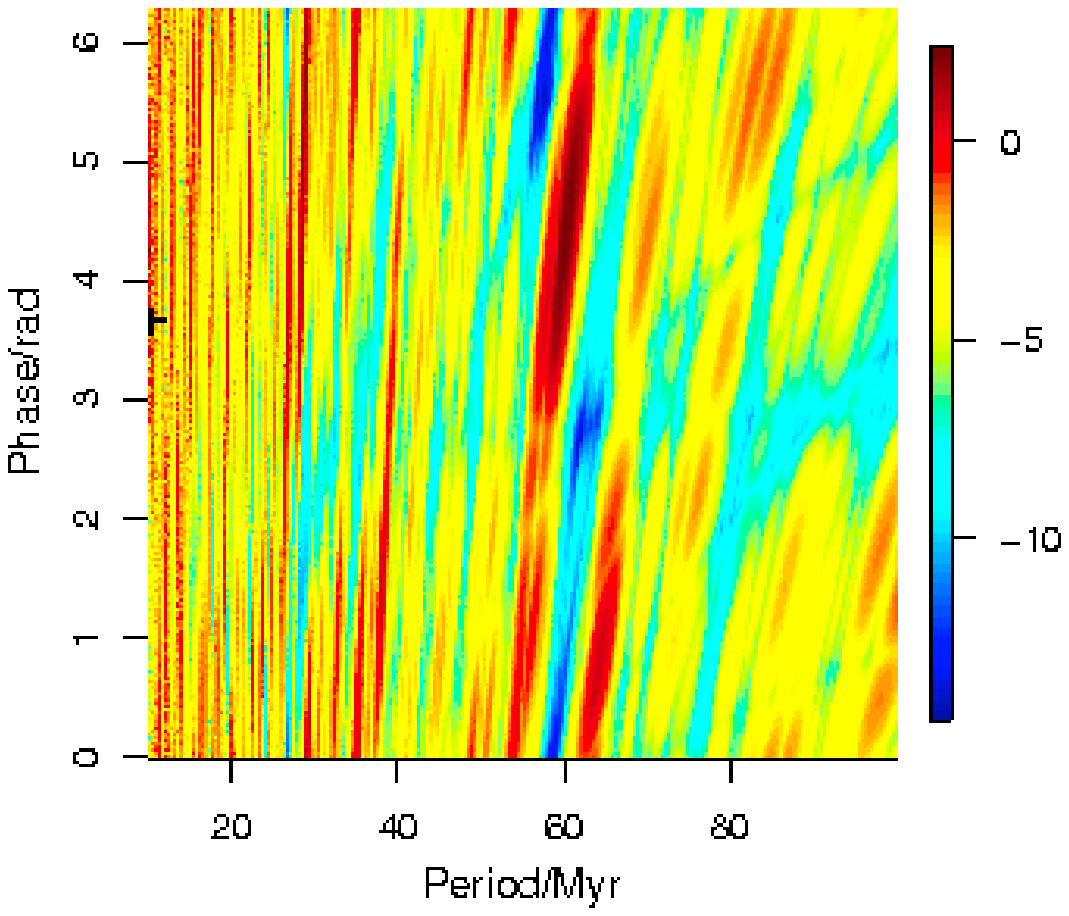}
\includegraphics[scale=0.7]{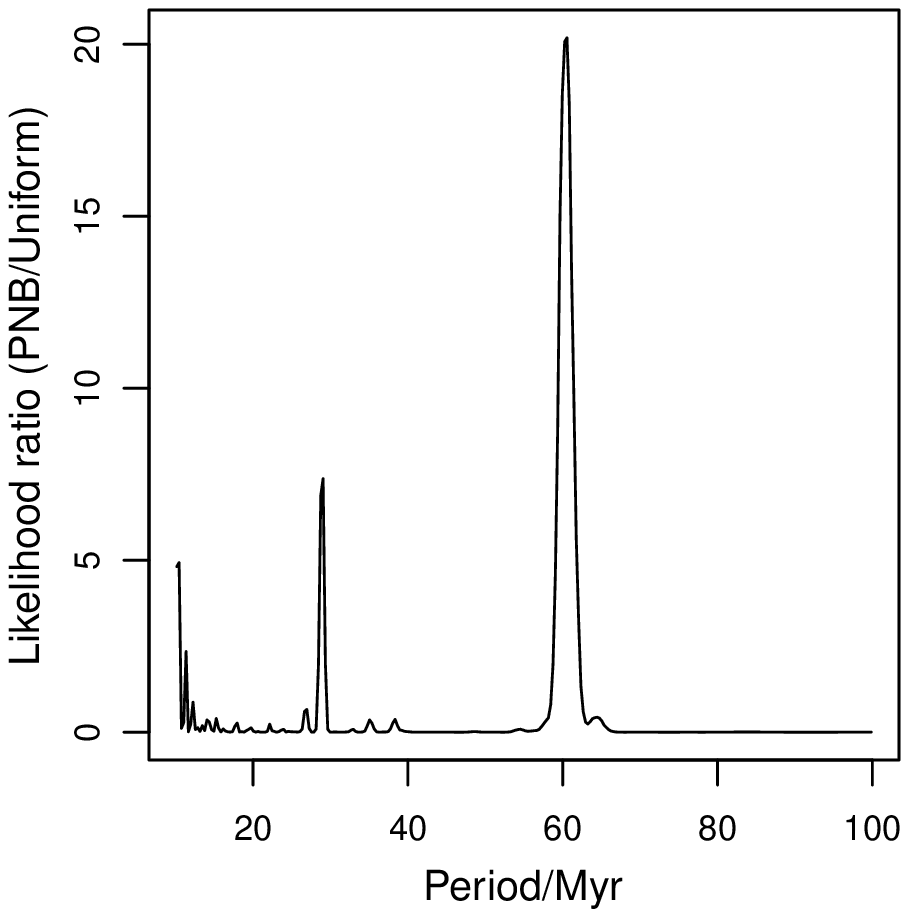}
\caption{Likelihood distribution for the PNB model on the B18 data set as
  a function of period and phase (upper panel) and period only (lower panel). In both panels we show the
likelihood ratio of the PNB to the Uniform model, on the left as log (base 10)  on a color scale.}
\label{fig:like-2D-pnb}
\end{figure}

Figure~\ref{fig:like-2D-pnb} shows how the likelihood varies over the
two-dimensional space formed by the two parameters, period and phase, of the
PNB model. There is significant variation. We see numerous local maxima, the
largest likelihoods being around $\{T/\text{Myr},~\beta/\text{rad}\} = \{60, 4.5\}$.  However, these maxima are rather narrow, so once the (much lower) likelihood in the other (equally plausible) regions are taken into account, the overall evidence for the model is not particularly high.  If we are interested in the variation of likelihood with just the period, then we can marginalize this diagram over phase, and plot with respect to period, thereby forming a (Bayesian) periodogram (lower panel). We see a clear peak around 60\,Myr. This is coincident with the period of $62\pm 3$\,Myr identified by \cite{rohde05}. It is tempting (but incorrect) to associate this peak value of the likelihood with the periodic model as a whole, and use it to claim a larger evidence for the periodic model. Certainly there is a degree of arbitrariness in the prior parameter distribution --- in this case a uniform distribution --- and narrowing this range around this peak would clearly increase the evidence.  For example, if we truncate the period range from its current value of [10,\,100]\,Myr to [50,\,80]\,Myr, then the Bayes factor relative to the Uniform model increases from 0.62 to 1.5. This is a rather modest increase, but we could increase it to a significantly high value with an even narrower prior. However, {\em we may not use the data to find the best fitting parameters and then claim that we should only consider the model near to these}. 
We would need some other reasoning or independent data for making such a
selection. (The \cite{rohde05} time series is not independent of B18 because
both are based on the same paleontological data.) We do not see how, a priori, we could limit the plausible periods of periodicity to something as narrow as 50--80\,Myr, let alone the much narrower range required to favor PNB significantly over other models. In the extreme limit of an infinitesimal region around the maximum likelihood, we end up doing model comparison using the maximum likelihood. Just out of interest, these values are shown in Table~\ref{tab:BF-MLR} and plotted in Figure~\ref{fig:BF-MLR}. If we were to use this (incorrect) metric, then PNB and some of its variants have significantly higher likelihood than the Uniform model and several of the other models (although barely more than a factor of ten above the random model, RB).  Another way of seeing why this is the wrong approach was already discussed in Section~\ref{sec:overview}: by focusing on the {\em best} fits we simply favor the more complex model. We could always define a more flexible model and so fit even better.

We labor this point because many of the claims for a periodicity in
biodiversity data have made use of a maximum likelihood approach (of which
$\chi^2$ is a special case) or something equivalent.  We must instead use the
evidence for model comparison. (Maximum likelihood may be used for estimating
the best parameters once we have established we have the best model.) If the periodic model were in fact the true one, then of course only one period and phase would be true. In that case the likelihood around these values would be so high as to result in a large evidence even when averaging over the broader parameter space (see simulations in Section~4.1 of \cite{bailer-jones11} for a demonstration).

Incidentally, the fact that we find a dominant period at all in the B18 data set is actually not that unlikely. The (Bayesian) periodogram of samples drawn at random from a uniform distribution often exhibits a period 
which has a likelihood larger than that of the true Uniform model (see Section~4.2 of \cite{bailer-jones11}). In other words, it is often possible to explain a random data set with {\em some} period, which is just a testament to how flexible the periodic model is.

\begin{figure*}
\centering
\includegraphics[scale=0.7]{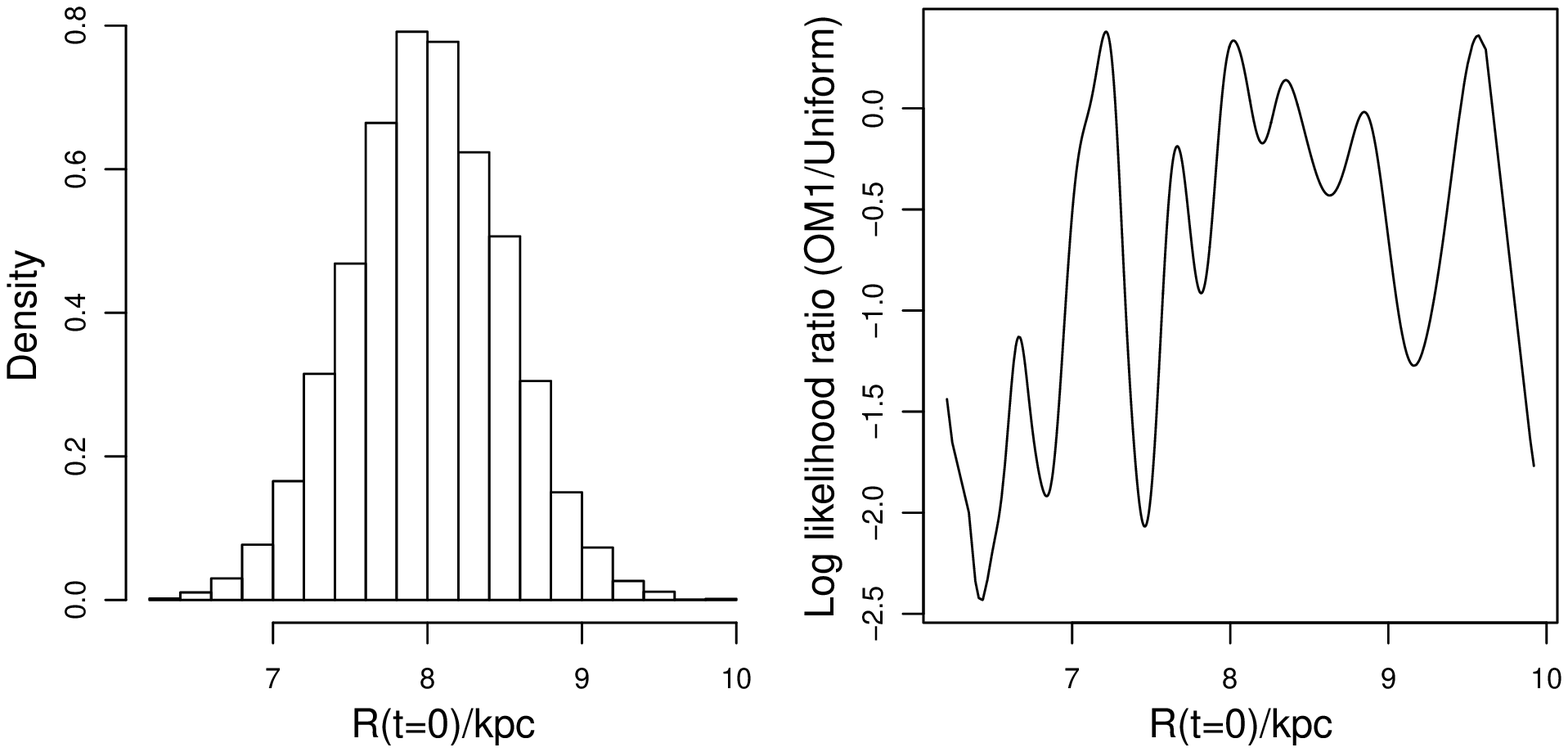}
\includegraphics[scale=0.7]{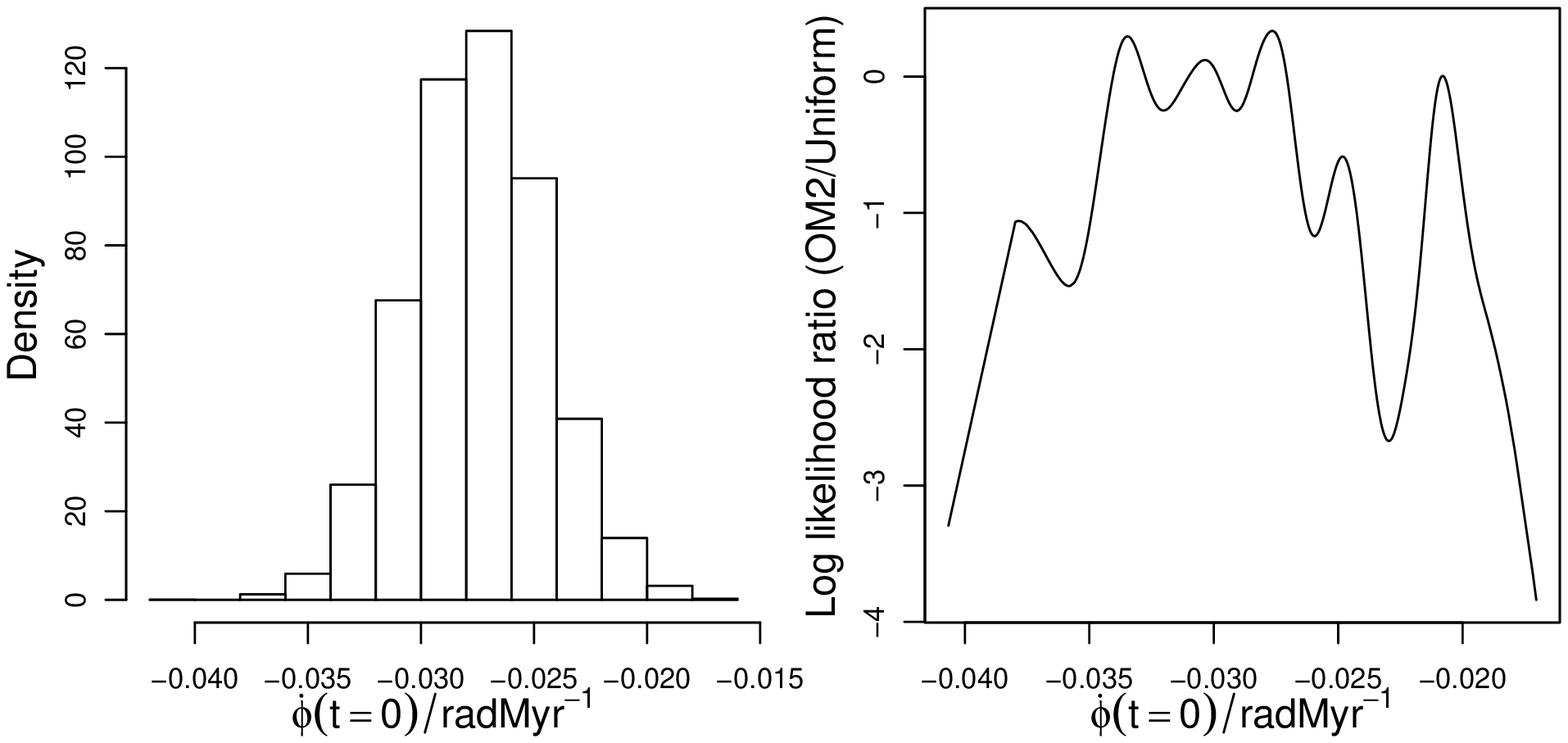}
\caption{The variation of initial conditions (left column) and corresponding
  variation of likelihood ratio relative to the Uniform model (right column) for the OM model on the B18 data set. The top row is for OM1, in which only $R(t=0)$ is varied. The bottom row is for OM2, in which only $\dot\phi(t=0)$ is varied.}
\label{fig:like-r-phidot-B18}
\end{figure*}

Moving on from the periodic model, we show in Figure~\ref{fig:like-r-phidot-B18}
the likelihood distribution for OM1 and OM2, i.e.\ where we vary the 
initial conditions $R(t=0)$ and $\dot\phi(t=0)$, respectively.
The likelihood ratio varies by a factor of up to $10^4$, but its absolute
value is never more than about two.  That is, no value of the intial
conditions gives a model much more favorable than the Uniform model, whereas
as some are far less favorable.  If we had lower uncertainties on these phase
space coordinates of the Sun, then we might be able to conclude something more definitive. For example, if $R(t=0)=7.5$\,kpc then the OM model would be even less favored.  We see a similar degree of variability for the other OM models and data sets listed in Table~\ref{tab:BF-MLR}.

We performed a similar analysis for the other model, but for the sake of space report only the maximum likelihood ratios in Table~\ref{tab:BF-MLR}.

For the B5 data set, the RNB model has the highest maximum likelihood ratio, yet its evidence was
the lowest. This indicates that while one particular instance of RNB fits the data well, overall it is a poor model.

For the RM and A08 data sets, the evidences are very similar for all models. This means that the data are not able to discriminate between these models very well: they are equally good (or bad).  However, the time series analysis model used here is not best suited to these data sets. These data can be better interpreted as valued time series, ones in which we have an extinction magnitude attached to each time (both, in general, with uncertainties), rather than a time variable probability of extinction.
\cite{bailer-jones12} has extended the present model in order to work with such data sets; the results of its application to RM and A08 will be reported in a future publication.

In summary, we find that for none of the data sets is any model particularly favored over the simple uniform one.
In particular, there is no evidence to suggest that the orbital model (with or without a background extinction level)
is a particularly good explanation of the data.

\subsection{Sensitivity test}\label{sec:sensitivity} 

The evidence is of course sensitive to the choice of prior parameter distribution, and often we have little reason to make a very specific choice. Here we test the sensitivity of the evidence to this, as well as to the parameters of the Galactic potential used in the orbital models, and to the age uncertainties for the discrete data sets.

\begin{figure}
\centering
\includegraphics[scale=0.7]{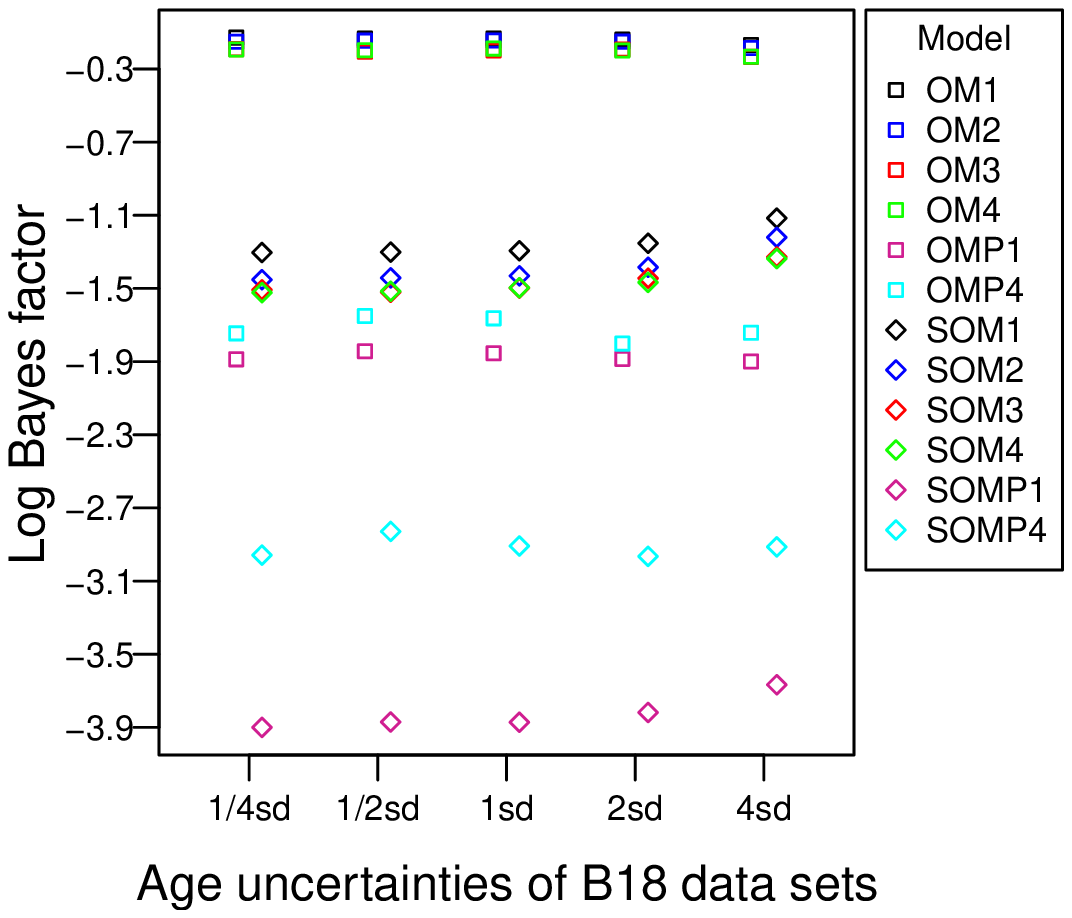}
\includegraphics[scale=0.7]{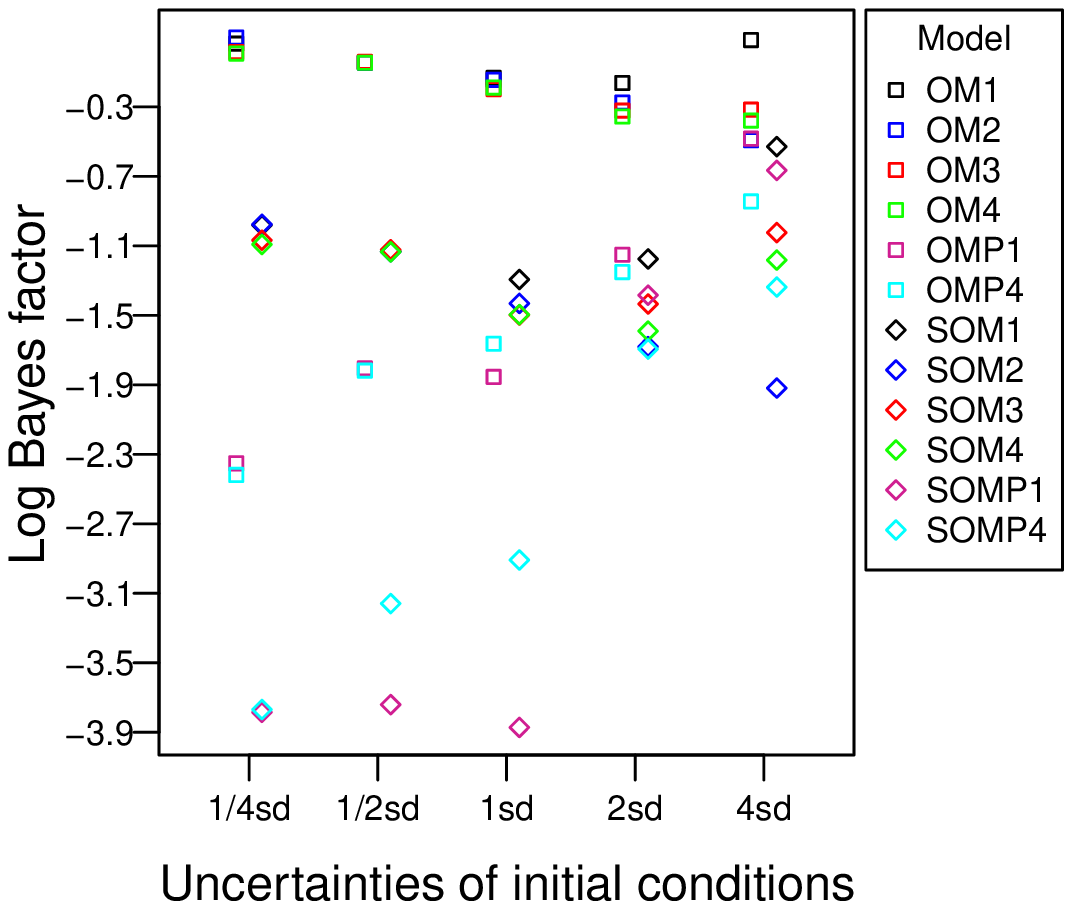}
\caption{Test of the sensitivity of the evidence to variations in the data age uncertainties and model priors.
Both panels plot (on the vertical axies) the values of the 
the Bayes factor of the varied OM(P) and SOM(P) models (relative to the
  Uniform model) for the B18 data sets. In the upper panel, the labels of the
  horizontal axis (Xsd) denote the varied B18 data sets, where Xsd means that we have
  multiplied the age uncertainties of the B18 time series by X for X=1/4, 1/2, 1, 2, 4.  
  In the lower panel, the labels of the horizontal axis (Xsd)
  denote variations of the uncertainties in (i.e.\ spread of) the initial conditions, where Xsd means
  multiplying the uncertainties of the initial conditions by a factor of X (X=1/4, 1/2, 1, 2, 4). 
  For each value of Xsd we have, for clarity, slightly offset in the horizontal direction the
  values for the OM(P) models (square points) from those for the SOM(P) models (diamond points).}
\label{fig:priortest}
\end{figure}

The age uncertainties in the discrete data are taken into account by the likelihood function. However, it is often difficult to estimate uncertainties, and we additionally made a plausible, but not unique, translation of the estimated duration of a stratigraphic substage in order to estimate uncertainty (which is the standard deviation of a Gaussian for each event; see Section~\ref{sec:datadiscrete}). To see how this affects our results, we scale the age uncertainties in the B18 data set by a constant factor of $1/4$, $1/2$, $2$, and $4$.  For each of these modified data sets we calculate the evidences for the models (S)OM1--4 and the Uniform model and recalculate the Bayes factors relative to the Uniform model.  These are plotted in the top panel of Figure~\ref{fig:priortest}: The Bayes factors change by just a few percent, so a precise age uncertainty is not necessary.

As a second test, we scale in the same way the uncertainties of the initial conditions of the orbital models (i.e.\ we change the width of the prior parameter distribution). The results are shown in the bottom panel of Figure~\ref{fig:priortest}.
The change in evidence for any particular model is larger than in the previous case, but in most cases less than a factor of 5 (except for the models including the spiral arms). Moreover, the absolute value of the Bayes factor remains below one. 

Our results are therefore also insensitive to considerable imprecision in the uncertainties in the phase space coordinates of the Sun. This (and the previous conclusion) is also true for the B5 data set.

As a third sensitivity test, we allow the number of simulated random events,
$N$, and the standard deviation of each event, $\sigma$, in the RNB and RB
models to vary. (Earlier we fixed $N=5$ when drawing models for the B5 data
set and $N=18$ for the B18 data set.) For the B18 data set, we find that a larger number of peaks or larger standard deviation in the RNB model produces a significantly larger Bayes factor (see Table~\ref{tab:sen-prior}), although it is still below unity. 
The RB model shows much less sensitivity to $N$ and $\sigma$.  We similarly recalculate the evidence for the other models in response to various perturbations of their priors, as also listed in Table~\ref{tab:sen-prior}.  The resulting Bayes factors do not change by more than a factor of 10 in any case, and often by much less.

\begin{table}
\caption{The Bayes factors (relative to the Uniform model) 
  on the B18 data set for models with priors varied. Each prior is varied
  individually (listed in the middle column) with the other fixed at their
  canonical values.}
\centering
\begin{tabular}{l|c|r}
\hline
\hline
models&varied prior&BF\\
\hline
\multirow{4}{*}{PNB}
&none&0.62\\
&$50<T<80$&1.5\\
&$10<T<200$&0.31\\
&$10<T<400$&0.15\\
\hline
\multirow{3}{*}{PB}
&none&0.80\\
&$B=1/2$&0.88\\
&$B=2$&0.97\\\hline
\multirow{4}{*}{QPM}
&none&0.85\\
&$0<A_Q<1/4$&0.62\\
&$0<A_Q<1$&0.54\\
&$100<T_Q<300$&0.61\\
&$100<T_Q<500$&0.58\\\hline
\multirow{5}{*}{RNB}
&none&0.00050\\
&$\sigma=5$~Myr&$3.7\times 10^{-11}$ \\ 
&$\sigma=20$~Myr&0.026\\
&$N=9$&0.0085\\
&$N=36$&0.29\\\hline
\multirow{5}{*}{RB}
&none&0.40\\
&$\sigma=5$~Myr&0.73\\
&$\sigma=20$~Myr&0.25\\
&$B=\frac{1}{2\sqrt{2\pi}\sigma}$&0.13\\
&$B=\frac{2}{\sqrt{2\pi}\sigma}$&0.45\\
&N=9&0.84\\
&N=36&0.20\\
\hline
\multirow{4}{*}{SP}
&none&0.019\\
&$-200<\lambda<0$&0.070\\
&$0<\lambda<200$&0.10\\
&$10<t_0<500$&0.047\\\hline
\multirow{3}{*}{SSP}
&none&0.18\\
&$10<T<200$&0.17\\
&$-200<\lambda<200$&0.37\\
&$10<t_0<500$&0.27\\
\hline
\end{tabular}
\label{tab:sen-prior}
\end{table}

Finally, we test the sensitivity to the Bayes factors to changes in the parameters
of the Galaxy model (canonical values listed in Table~\ref{tab:modelpar}).
The results are shown in Table~\ref{tab:sen-model-par}) for the B18 data set. If we double the mass of the
halo, for example, then the evidence for the OM and SOM models changes by no more than a factor of three.
Some other changes produce smaller effects, some larger, but not more than by a factor of 5 (and note that a change in a factor of two of the scale lengths is beyond what is consistent with observed data). 
Changes in the parameters of models which include spiral arms (the OMP and SOMP models) can produce
larger changes in the Bayes factors.
However, most significantly,
none of these changes produce a Bayes factor greater than one. In other words, none of these changes result in the orbital model becoming a better explanation for the paleontological than the Uniform model. 

\begin{table*}
\centering
\small
\caption{The Bayes factors on the B18 data set for orbital models with varied Galaxy parameters.}
\begin{tabular}{l*{12}{c}}
\hline
\hline
variation          &OM1  &OM2  &OM3 &OM4 &OMP1   &OMP4  &SOM1  &SOM2 &SOM3  &SOM4   &SOMP1    &SOMP4\\
\hline
none          &0.74 &0.72 &0.63&0.65&0.014  &0.022 &0.051 &0.037 &0.032  &0.032&1.3$\times$10$^{-4}$&1.2$\times$10$^{-3}$\\
$2M_b$          &0.58 &0.45 &0.54&0.55&0.011  &0.018 &0.043 &0.011 &0.027 &0.028 &5.8$\times$10$^{-4}$&3.8$\times$10$^{-4}$\\
$1/2M_b$        &0.78 &0.85 &0.67&0.67&0.0066 &0.021 &0.040 &0.045 &0.030 &0.030 &1.4$\times$10$^{-4}$&1.9$\times$10$^{-3}$\\
$2b_b$          &0.74 &0.72 &0.64&0.63&0.0036 &0.013 &0.055 &0.042 &0.033 &0.032 &4.6$\times$10$^{-5}$&6.0$\times$10$^{-4}$\\
$1/2b_b$        &0.74 &0.73 &0.65&0.63&0.019  &0.025 &0.051 &0.037 &0.032 &0.030 &2.0$\times$10$^{-4}$&1.5$\times$10$^{-3}$\\
$2 M_h$         &0.60 &0.68 &0.57&0.57&0.00014&0.0031&0.080 &0.069 &0.087 &0.083 &4.6$\times$10$^{-6}$&9.4$\times$10$^{-5}$\\
$1/2 M_h$       &0.81 &0.68 &0.66&0.66&0.011  &0.014 &0.20  &0.059 &0.15  &0.15  &2.3$\times$10$^{-4}$&2.8$\times$10$^{-4}$\\
$2 b_h$         &0.81 &0.52 &0.66&0.65&0.011  &0.0061&0.19  &0.027 &0.15  &0.15  &1.1$\times$10$^{-3}$&3.5$\times$10$^{-4}$\\
$1/2 b_h$       &0.073&0.048&0.11&0.11&0.024  &0.062 &0.0046&0.0012&0.010 &0.010 &7.5$\times$10$^{-3}$&1.4$\times$10$^{-2}$\\
$2 M_d$         &0.21 &0.063&0.33&0.32&0.035  &0.058 &0.0047&0.0046&0.027 &0.024 &2.2$\times$10$^{-5}$&7.6$\times$10$^{-4}$\\
$1/2 M_d$       &0.59 &0.56 &0.49&0.48&0.0031 &0.0037&0.0073&0.0018&0.0086&0.0088&3.3$\times$10$^{-4}$&9.6$\times$10$^{-4}$\\
$2 a_d$         &0.86 &0.89 &0.76&0.76&0.13   &0.12  &0.10  &0.088 &0.069 &0.067 &6.3$\times$10$^{-3}$&7.0$\times$10$^{-3}$\\
$1/2 a_d$       &0.63 &0.66 &0.55&0.55&0.021  &0.019 &0.015 &0.011 &0.022 &0.026 &2.5$\times$10$^{-3}$&9.1$\times$10$^{-4}$\\
$2 b_d$         &1.1  &1.1  &0.93&0.93&0.0056 &0.020 &0.028 &0.030 &0.027 &0.025 &1.3$\times$10$^{-4}$&4.2$\times$10$^{-3}$\\
$1/2 b_d$       &0.72 &0.87 &0.60&0.56&0.00085&0.0073&0.15  &0.17  &0.10  &0.090 &1.3$\times$10$^{-4}$&1.5$\times$10$^{-3}$\\
\hline
\end{tabular}
\label{tab:sen-model-par}
\end{table*}

In summary, we find that the evidences for most models are not particularly sensitive to the age uncertainties, Galaxy model parameters, or reasonable changes to the prior parameter distributions.

\subsection{Testing the discriminative power}\label{sec:power}

Here we investigate how well our analysis can discriminate between models, by using simulated data which have been drawn from one of these models. Given sufficient data, such a discrimination will always be possible to some threshold Bayes factor, but here we are interested in the case where the data have similar properties (in particular, sparsity) to the real data we have been using.

We investigate this by simulating a number of of time series from the RNB, OM1, and PNB models (which we will refer to as ``generative models'' when used in this way, to distinguish from their use to calculate the evidence on given data). For each generative model, we fix the parameters to certain values, then sample 18 events from the resulting $P(t_j|\theta, M)$ to give a simulated time series, to which we then attach the measured age uncertainties. We generate ten time series in this way (and below we average the Bayes factors over these and report that). For the OM1 and PNB generative models we repeat this at ten different values of the solar initial radius parameter (OM1) or period parameter (PNB) in the generative model.  (RNB has no parameters). We repeat the whole process a second time but using simulated age uncertainties drawn from a log normal distribution with standard deviation and mean calculated from the measured age uncertainties. We finally repeat the process a third time for the OM1 and PNB generative model, but now drawing data to have the same time sampling as our continuous data sets (for which age uncertainties are not used; see section~\ref{sec:continuous}). (We do not do this for the RNB model as it predicts discrete events.)

For each simulated data set we calculate Bayes factors for the RNB, OM1, and PNB models relative to the Uniform model. For the data generated from the RNB model (with age uncertainties taken from the data), the Bayes factors for the three models are as follows: 0.18 for RNB; 0.57 for OM1; 0.52 for PNB. (We get almost identical values when the age uncertainties were drawn at random). Thus no model -- not even the true one -- is favored over the Uniform model (although none is significantly rejected either). This is not that surprising, however, because with only 18 events, and with the evidence effectively averaging the predicted times of events from the RNB model over all time, the Uniform and RNB models end up with similar predictive power. This is unavoidable, because with the RNB model we cannot decide in advance where the events are: we must average over all possibilities.

\begin{figure*}
  \centering \includegraphics[scale=0.5]{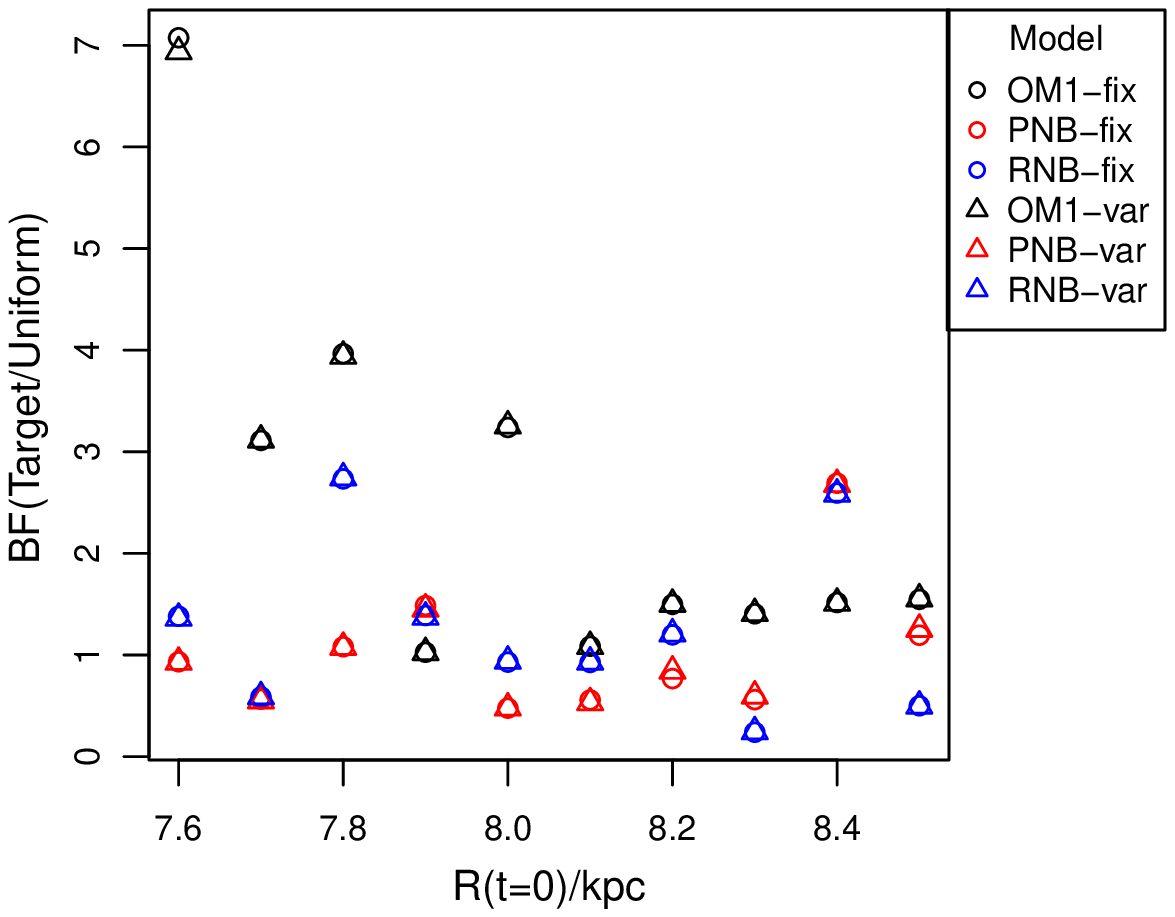} \includegraphics[scale=0.5]{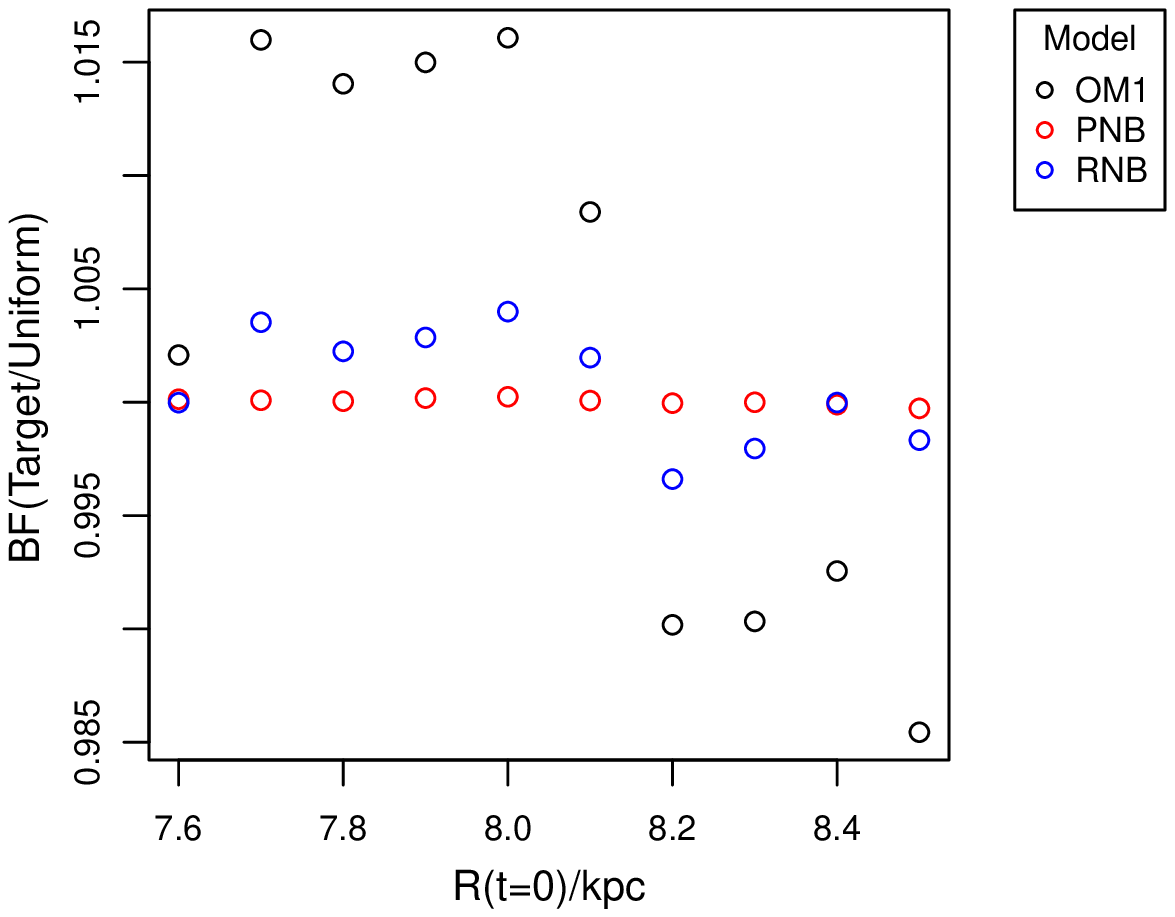} \includegraphics[scale=0.5]{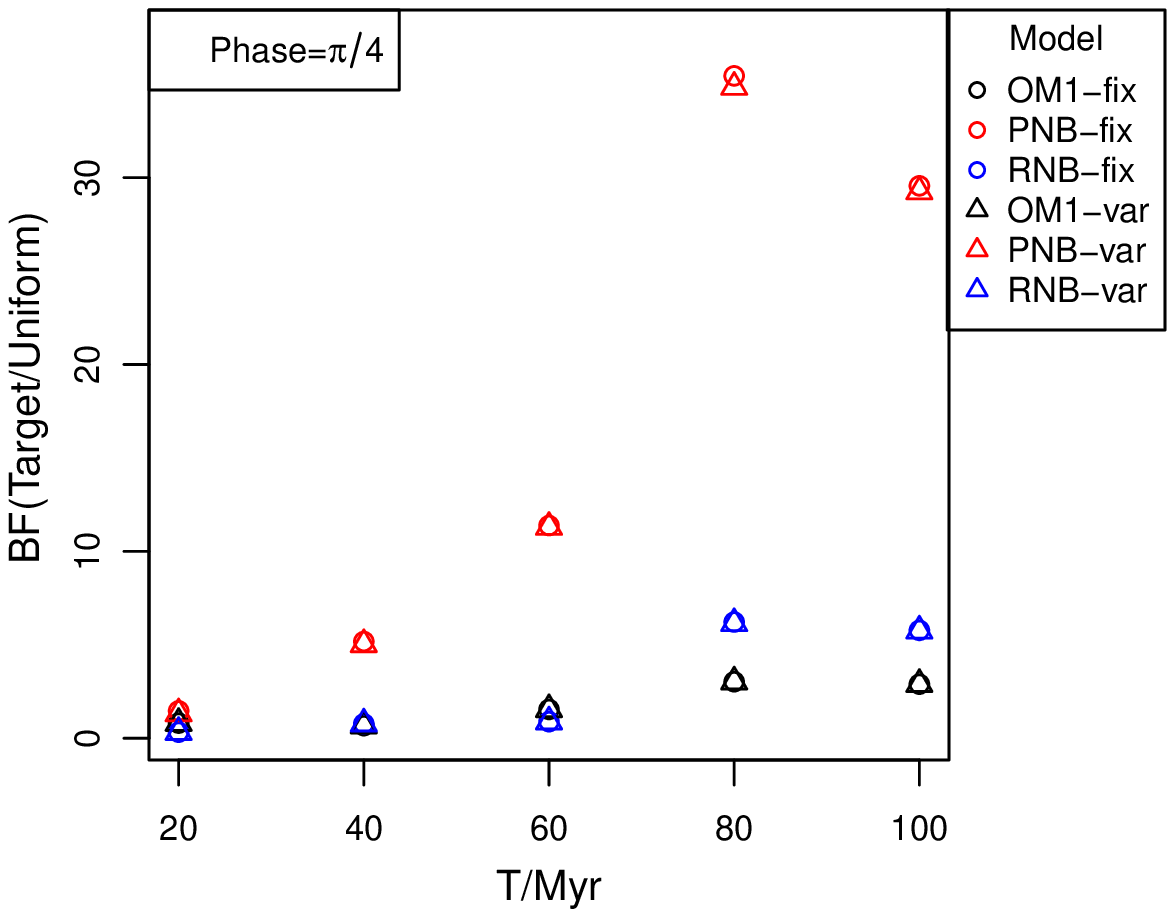} \includegraphics[scale=0.5]{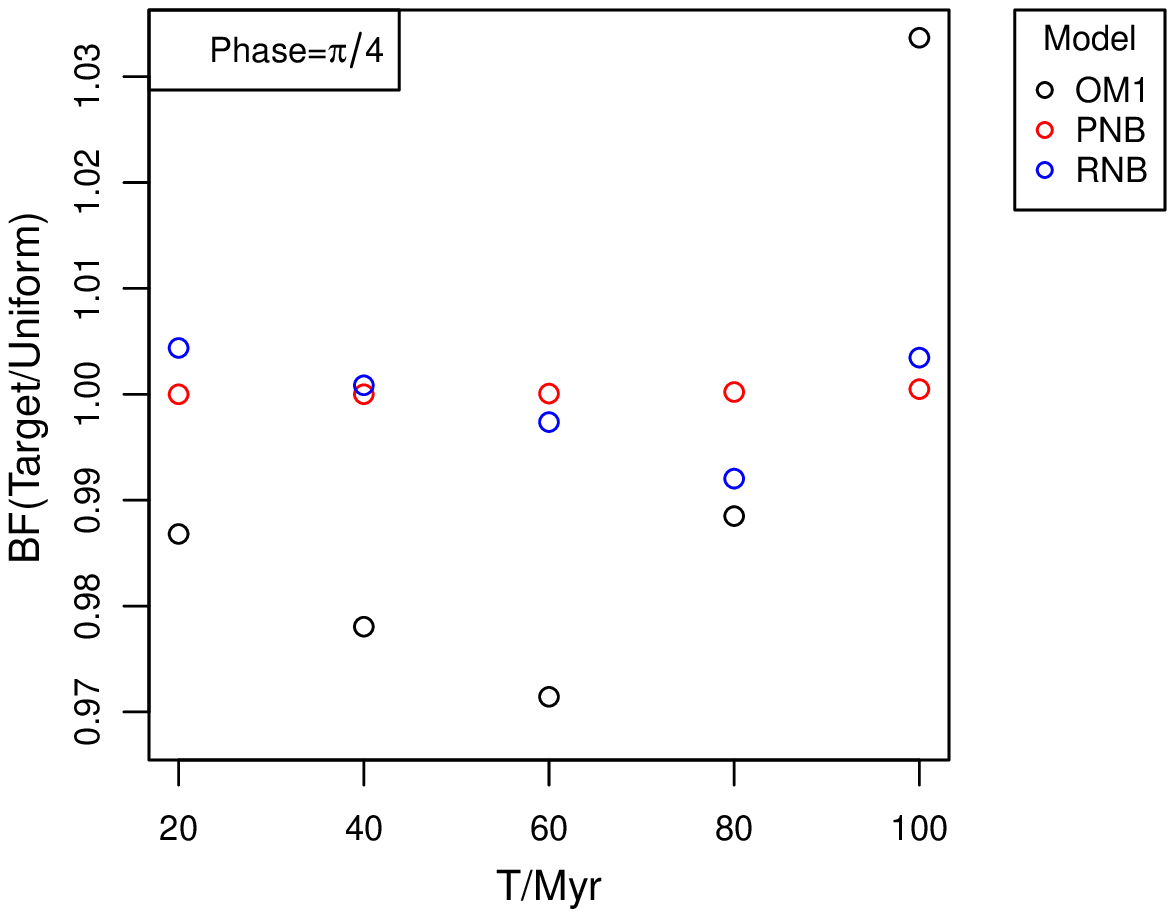} \includegraphics[scale=0.5]{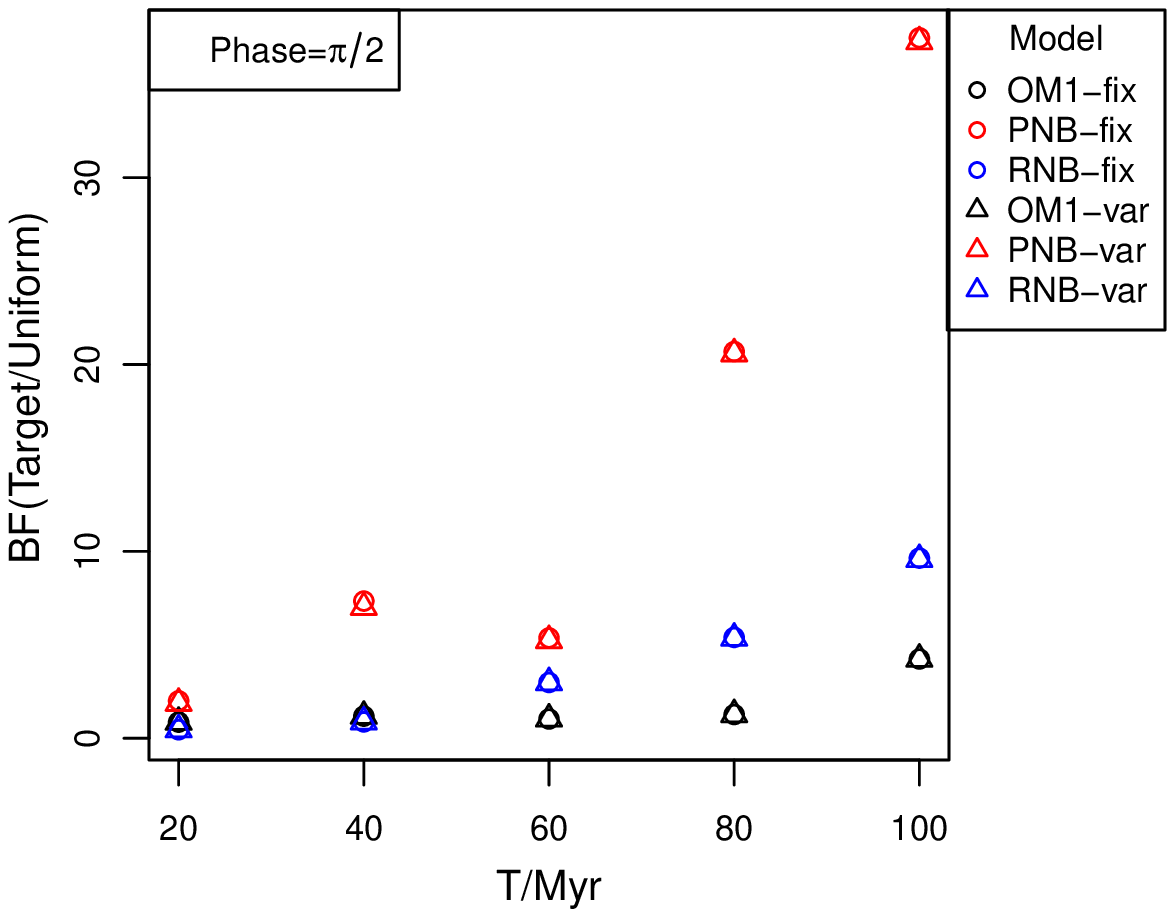} \includegraphics[scale=0.5]{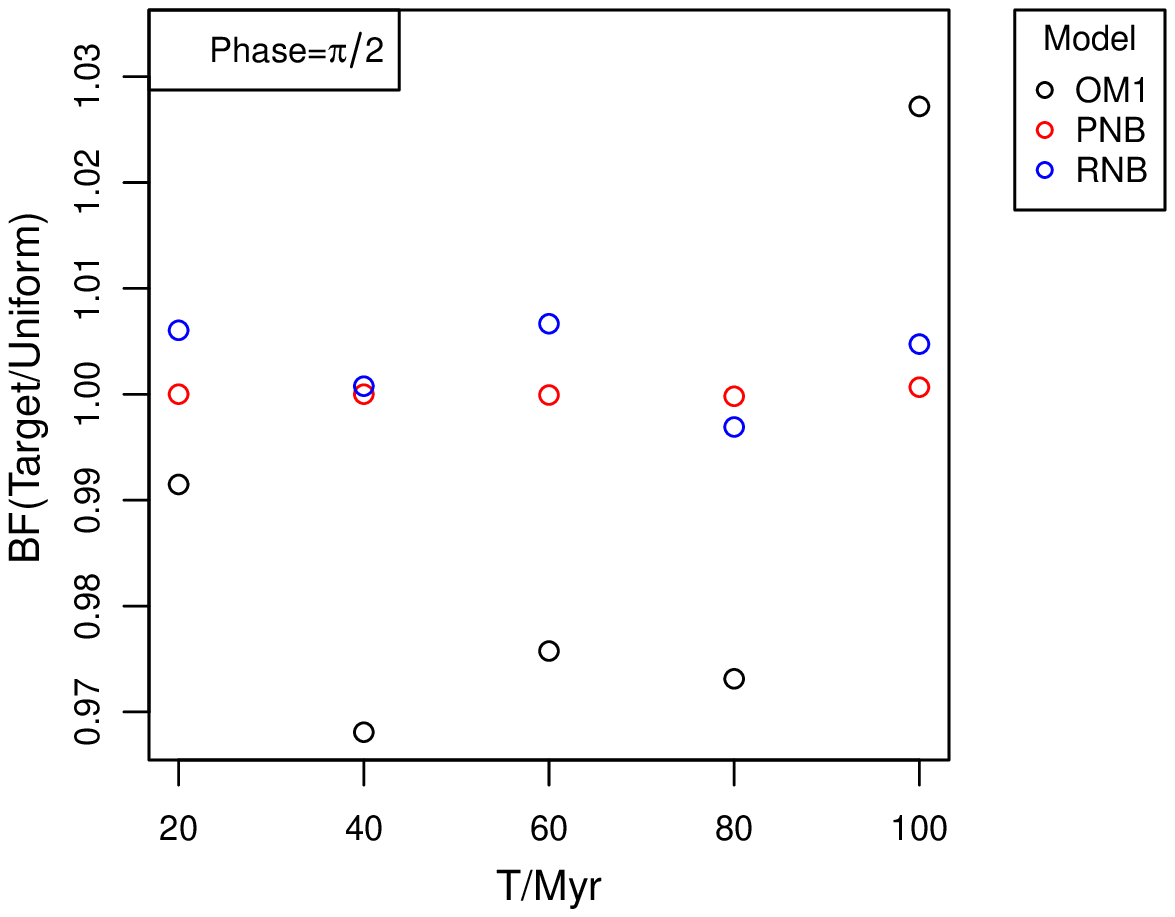} \caption{Bayes
    factors for various models computed on simulated data sets. The horizontal
    axis in each panel indicates the value of the parameter in the model used
    to generate these data sets (all other parameters are kept fixed). The
    left panels are for discrete data sets and the right ones for continuous
    data sets.  The top row is for data drawn from the OM1 model, and the
    bottom two rows are for data drawn from the PNB model (for two different values of that generative model's phase parameter, $\pi/4$ and $\pi/2$). The models for which the Bayes factors have been computed are shown in different symbols defined in the plot: OM1, PNB and RNB. The suffices ``fix'' and ``var'' indicate age uncertainties for the discrete data either taken from the real data or drawn from a log normal distribution (respectively). }
\label{fig:sim-data-comp}
\end{figure*}

The results for applying the models to the data generated from the OM1 and PNB models are shown in 
Fig.~\ref{fig:sim-data-comp}, where the horizontal axis shows how the Bayes factor varies with the one parameter which is varied in these generative models.  The top row shows the results for data drawn from the OM1 model, for the discrete (B18-like) data (left) and the continuous data (right). We see that the (true) OM1 model is not significantly favored in either case (Bayes factor always less than 10), although not disfavored either (Bayes factor more than 0.1). In particular, the continuous data show no discriminative power.

The lower two rows show conceptually the same thing, but now for data drawn from the PNB model for two different values of the phase parameter (the two rows). Here we see that, at least for longer periods, the PNB model is generally correctly identified (on the basis of a large Bayes factor), when using the discrete data sets. Yet the continuous data still show no discriminative power.

This difference between discrete and continuous data sets is not unexpected.
In the former, the likelihood is a product of the likelihood for many events, each of which is the convolution
of the event with the model. In the latter, the likelihood is just the result of a single convolution 
of a continuous model over continuous data. This is not the best approach for modeling continuous data. A better choice is the recently developed continuous time series modeling method described in 
\cite{bailer-jones12}, which will be used in future work.

Clearly one could perform many more tests with more simulated time series, varying different parameters in the generative models and with different permutations of the values of the fixed parameters. No doubt there are parts of parameter space where some models are favored over others, in particular if we adopted more informative priors. Thus while these results on simulated data give some check on the discriminative power of the method and data, they should not be over-interpreted to say anything too general. Nonetheless, the tests we have done confirm what we concluded based on the analysis of the real data. Specifically, while our analysis of the real data does not allow us to claim evidence in favor of the orbital-based models, it also cannot rule out these models. This is due partly to the lack of predictive power of the data, and partly to the large flexibility (or broad prior parameter space) of the models. Better constraints on the solar orbit would help reduce the latter.

\section{Summary and conclusions}\label{sec:conclusion} 

We have used a Bayesian model comparison method to examine how well different time series models explain the variation of biodiversity over the Phanerozoic eon (the past 550\,Myr). One class of models is derived from the orbit of the Sun around the Galaxy, which we reconstructed from a model of the Galactic mass distribution. Our model comparison takes into account uncertainties in the data as well as uncertainties in the reconstructed path of the Sun. We have compared the evidence for this model with that of various other reference models of no particular causal origin. All models are stochastic in the sense that they predict only the time variation of the extinction probability, rather than the exact magnitude of the extinction rate or the times of mass extinctions.

As part of this analysis we investigated the properties of plausible solar orbits (i.e.\ those consistent with the  accuracy of the present phase space coordinates of the Sun). We find that the majority of orbits have a motion perpendicular to the Galactic plane which is not far from periodic, although a precise period cannot be inferred due to the uncertainties in the present solar phase space coordinates as well as the exact mass distribution (gravitational potential) of the Galaxy. Thus any claims which try to link a variation in geological biodiversity, cratering or climate records to this solar motion must consider the motion as quasi-periodic rather than strictly periodic.

In contrast, only about half of the simulated orbits showed periodic spiral arm crossings, even for a very simple, rigidly rotation arm model. Indeed, many of the orbits did not encounter the spiral arms more than once. It should be noted that the shape and pattern speed of the spiral arms is poorly known (and they may not even be long-lived), so any claims of a causal connection between spiral arm passages and terrestrial conditions should be treated with due skepticism.
 
We have shown how the evidence (marginal likelihood) should be used to do model comparison, as opposed to selecting the model which gives the best single fit. The reason is that an arbitrarily complex model can always be tuned to fit the data arbitrarily well, yet that does not make it a good model.  By averaging the likelihood over the parameter space, the evidence uses the rules of probability to trade off the quality of the fit with the model plausibility in a quantitative fashion. Of the models investigated, we do not claim any one of them to be ``true''. Indeed, no model is exactly true in reality.  All we can hope to do is to find the best of the ones tested so far.

We find that none of the models tested --- including periodic, quasi-periodic and orbital-based --- explain the discrete data sets better than a Uniform model. In other words, the time distribution of mass extinction events is consistent with being randomly distributed in time. There is no need to resort to anything more exotic.

The Uniform model is also no worse than other models for the continuous data sets. This does not mean that we believe the extinction rate has been constant over the Phanerozoic, but rather that none of the other (more complex) models is {\em significantly} better. Assuming the variations in extinction seen in Figure~\ref{fig:data} are true --- something we have no reason to doubt --- then this tells us that there must be other models, not yet tested, which could explain the data better.  This will be investigated in future work using a model more suited for these types of time series.

We found in particular that the orbital-based extinction model is not favored by the data. This conclusion is robust to changes in the parameters of the Galaxy model and to the magnitude of the uncertainties of both the solar phase space coordinates and the ages of the extinction events.  
On the other hand, our analysis of simulated data showed that even if the
orbital model were the true one, our analysis could not have identified it
with either the discrete or (in particular) the continuous data sets. This
ultimately comes down to a combination of a lack of discriminative power in the data, plus a large flexibility (or prior parameter space) in the models.
Of course, if the orbit of the Sun could be much better determined then it is possible that this model would then be more --- or less --- favored by our analysis.  We remind the reader that our orbital model adopted an extinction mechanism in which the extinction rate is proportional to the integrated ``flux'' (of a non-specified type) from nearby stars. A radical change in this mechanism would of course correspond to a quite different model, which could give different results.  Thus we do not claim that the solar motion plays no part in terrestrial extinction, nor that astronomical mechanisms are irrelevant.

Indeed, it is quite plausible that the biological extinction rate has been
affected by many factors, and so any attempt to connect them solely to the
solar motion, or indeed to any simple analytic model, is doomed from the
start. We have addressed this to some extent by including compound models and
the semi-orbital model, but clearly one could do more. However, given the
present uncertainties of the reconstructed solar orbit, it seems unlikely that
one could draw a strong conclusion on the positive relevance of the solar
orbit on the basis of current geological data.  This, indeed, is the main
conclusion of this work, plus the confirmation that periodic models are not a
good (or necessary) explanation of the biodiversity variation.  There is some
hope that, in the future, results from the {\it Gaia} survey of the Galaxy (e.g., \cite{lindegren08}) will improve our knowledge of the Galactic potential, spiral arms and inferred solar orbit, to the extent that this study can be repeated to give conclusions of greater certainty.

\section*{Acknowledgments}

We thank Glenn Van de Ven for discussions on the orbital
modeling as well as the referee for useful comments on the manuscript. This
work has been carried out as part of the Gaia Research for European Astronomy
Training (GREAT-ITN) network. The research leading to these results has received funding from the European Union Seventh Framework Programme ([FP7/2007-2013] under grant agreement No. 264895.


\begin{thebibliography}{79}
\expandafter\ifx\csname natexlab\endcsname\relax\def\natexlab#1{#1}\fi

\bibitem[{{Alroy}(1996)}]{alroy96}
{Alroy}, J. 1996, Palaeogeography, Palaeoclimatology, Palaeoecology, 127, 285 ,
  285

\bibitem[{{Alroy}(2008)}]{alroy08}
---. 2008, Proceedings of the National Academy of Science, 105, 11536, 11536

\bibitem[{{Alroy}(2010)}]{alroy10}
---. 2010, Palaeontology, 53, 1211–1235, 1211–1235

\bibitem[{{Alroy} {et~al.}(2001){Alroy}, {Marshall}, {Bambach}, {Bezusko},
  {Foote}, {F{\"u}rsich}, {Hansen}, {Holland}, {Ivany}, {Jablonski}, {Jacobs},
  {Jones}, {Kosnik}, {Lidgard}, {Low}, {Miller}, {Novack-Gottshall},
  {Olszewski}, {Patzkowsky}, {Raup}, {Roy}, {Sepkoski}, {Sommers}, {Wagner}, \&
  {Webber}}]{alroy01}
{Alroy}, J., {Marshall}, C.~R., {Bambach}, R.~K., {et~al.} 2001, Proceedings of
  the National Academy of Science, 98, 6261, 6261

\bibitem[{{Alroy} {et~al.}(2008){Alroy}, {Aberhan}, {Bottjer}, {Foote},
  {F{\"u}rsich}, {Harries}, {Hendy}, {Holland}, {Ivany}, {Kiessling}, {Kosnik},
  {Marshall}, {McGowan}, {Miller}, {Olszewski}, {Patzkowsky}, {Peters},
  {Villier}, {Wagner}, {Bonuso}, {Borkow}, {Brenneis}, {Clapham}, {Fall},
  {Ferguson}, {Hanson}, {Krug}, {Layou}, {Leckey}, {N{\"u}rnberg}, {Powers},
  {Sessa}, {Simpson}, {Toma{\v s}ov{\'y}ch}, \& {Visaggi}}]{alroy08sci}
{Alroy}, J., {Aberhan}, M., {Bottjer}, D.~J., {et~al.} 2008, Science, 321, 97,
  97

\bibitem[{{Alvarez} {et~al.}(1980){Alvarez}, {Alvarez}, {Asaro}, \&
  {Michel}}]{alvarez80}
{Alvarez}, L.~W., {Alvarez}, W., {Asaro}, F., \& {Michel}, H.~V. 1980, Science,
  208, 1095, 1095

\bibitem[{{Alvarez} \& {Muller}(1984)}]{alvarez84}
{Alvarez}, W., \& {Muller}, R.~A. 1984, \nat, 308, 718, 718

\bibitem[{Atri \& Melott(2013)}]{atri12}
Atri, D., \& Melott, A.~L. 2013, Astroparticle Physics, ,

\bibitem[{{Bahcall} \& {Bahcall}(1985)}]{bahcall85}
{Bahcall}, J.~N., \& {Bahcall}, S. 1985, \nat, 316, 706, 706

\bibitem[{{Bailer-Jones}(2009)}]{bailer-jones09}
{Bailer-Jones}, C.~A.~L. 2009, International Journal of Astrobiology, 8, 213,
  213

\bibitem[{{Bailer-Jones}(2011{\natexlab{a}})}]{bailer-jones11}
---. 2011{\natexlab{a}}, \mnras, 416, 1163, 1163

\bibitem[{{Bailer-Jones}(2011{\natexlab{b}})}]{bailer-jones11-err}
---. 2011{\natexlab{b}}, \mnras, 418, 2111, 2111

\bibitem[{{Bailer-Jones}(2012)}]{bailer-jones12}
---. 2012, \aap, 546, A89, A89

\bibitem[{{Bambach}(2006)}]{bambach06}
{Bambach}, R.~K. 2006, Annual Review of Earth and Planetary Sciences, 34, 127,
  127

\bibitem[{{Bambach Richard K.} {et~al.}(2004){Bambach Richard K.}, {Knoll
  Andrew H.}, \& {Wang Steve C.}}]{bambach04}
{Bambach Richard K.}, {Knoll Andrew H.}, \& {Wang Steve C.} 2004, Paleobiology,
  30, 522–542, 522–542, doi:
  10.1666/0094-8373(2004)030<0522:OEAMDO>2.0.CO;2

\bibitem[{Barnosky(2001)}]{barnosky01}
Barnosky, A.~D. 2001, Journal of Vertebrate Paleontology, 21, 172, 172

\bibitem[{Benton(2009)}]{benton09}
Benton, M.~J. 2009, Science, 323, 728, 728

\bibitem[{Carslaw {et~al.}(2002)Carslaw, Harrison, \& Kirkby}]{carslaw02}
Carslaw, K.~S., Harrison, R.~G., \& Kirkby, J. 2002, Science, 298, 1732, 1732

\bibitem[{{Cox} \& {G{\'o}mez}(2002)}]{cox02}
{Cox}, D.~P., \& {G{\'o}mez}, G.~C. 2002, \apjs, 142, 261, 261

\bibitem[{{Crowley} \& {North}(1988)}]{crowley88}
{Crowley}, T.~J., \& {North}, G.~R. 1988, Science, 240, 996, 996

\bibitem[{{Davis} {et~al.}(1984){Davis}, {Hut}, \& {Muller}}]{davis84}
{Davis}, M., {Hut}, P., \& {Muller}, R.~A. 1984, \nat, 308, 715, 715

\bibitem[{{Dehnen} \& {Binney}(1998)}]{dehnen98b}
{Dehnen}, W., \& {Binney}, J.~J. 1998, \mnras, 298, 387, 387

\bibitem[{Domainko {et~al.}(2013)Domainko, Bailer-Jones, \& Feng}]{domainko13}
Domainko, W., Bailer-Jones, C. A.~L., \& Feng, F. 2013, Monthly Notices of the
  Royal Astronomical Society,
  http://mnras.oxfordjournals.org/content/early/2013/04/10/mnras.stt455.full.pdf+html

\bibitem[{{Drimmel}(2000)}]{drimmel00}
{Drimmel}, R. 2000, \aap, 358, L13, L13

\bibitem[{{Eisenhauer} {et~al.}(2003){Eisenhauer}, {Sch{\"o}del}, {Genzel},
  {Ott}, {Tecza}, {Abuter}, {Eckart}, \& {Alexander}}]{eisenhauer03}
{Eisenhauer}, F., {Sch{\"o}del}, R., {Genzel}, R., {et~al.} 2003, \apjl, 597,
  L121, L121

\bibitem[{{Ellis} \& {Schramm}(1995)}]{ellis95}
{Ellis}, J., \& {Schramm}, D.~N. 1995, Proceedings of the National Academy of
  Science, 92, 235, 235

\bibitem[{{Feulner}(2009)}]{feulner09}
{Feulner}, G. 2009, International Journal of Astrobiology, 8, 207, 207

\bibitem[{{Garc{\'{\i}}a-S{\'a}nchez}
  {et~al.}(2001){Garc{\'{\i}}a-S{\'a}nchez}, {Weissman}, {Preston}, {Jones},
  {Lestrade}, {Latham}, {Stefanik}, \& {Paredes}}]{sanchez01}
{Garc{\'{\i}}a-S{\'a}nchez}, J., {Weissman}, P.~R., {Preston}, R.~A., {et~al.}
  2001, \aap, 379, 634, 634

\bibitem[{{George} \& {Hao}(2009)}]{sugihara09}
{George}, S., \& {Hao}, Y. 2009, Nature, 458, 979, 979, 10.1038/458979a

\bibitem[{{Gies} \& {Helsel}(2005)}]{gies05}
{Gies}, D.~R., \& {Helsel}, J.~W. 2005, \apj, 626, 844, 844

\bibitem[{Glen(1994)}]{glen94}
Glen, E. 1994, The Mass-Extinction Debates: How Science Works in a Crisis

\bibitem[{{Hallam}(1989)}]{hallam89}
{Hallam}, A. 1989, Royal Society of London Philosophical Transactions Series B,
  325, 437, 437

\bibitem[{{Hays} {et~al.}(1976){Hays}, {Imbrie}, \& {Shackleton}}]{hays76}
{Hays}, J.~D., {Imbrie}, J., \& {Shackleton}, N.~J. 1976, Science, 194, 1121,
  1121

\bibitem[{{Holland}(2012)}]{holland12}
{Holland}, S.~M. 2012, Paleobiology, 38, 205–217, 205–217, doi:
  10.1666/11030.1

\bibitem[{Jeffreys(1961)}]{jeffreys61}
Jeffreys, H. 1961, Theory of probability, 3rd edn., The international series of
  monographs on physics (Oxford: Clarendon Press),

\bibitem[{Kass \& Raftery(1995)}]{kass95}
Kass, R.~E., \& Raftery, A.~E. 1995, Journal of the American Statistical
  Association, 90, 773–795, 773–795

\bibitem[{{Kirkby}(2007)}]{kirkby08}
{Kirkby}, J. 2007, Surveys in Geophysics, 28, 333, 333

\bibitem[{{Kitchell} \& {Pena}(1984)}]{kitchell84}
{Kitchell}, J.~A., \& {Pena}, D. 1984, Science, 226, 689, 689

\bibitem[{{Koyama} {et~al.}(1995){Koyama}, {Petre}, {Gotthelf}, {Hwang},
  {Matsuura}, {Ozaki}, \& {Holt}}]{koyama95}
{Koyama}, K., {Petre}, R., {Gotthelf}, E.~V., {et~al.} 1995, \nat, 378, 255,
  255

\bibitem[{{Leitch} \& {Vasisht}(1998)}]{leitch98}
{Leitch}, E.~M., \& {Vasisht}, G. 1998, New Astronomy, 3, 51, 51

\bibitem[{{Lindegren} {et~al.}(2008){Lindegren}, {Babusiaux}, {Bailer-Jones},
  {Bastian}, {Brown}, {Cropper}, {H{\o}g}, {Jordi}, {Katz}, {van Leeuwen},
  {Luri}, {Mignard}, {de Bruijne}, \& {Prusti}}]{lindegren08}
{Lindegren}, L., {Babusiaux}, C., {Bailer-Jones}, C., {et~al.} 2008, in IAU
  Symposium, Vol. 248, IAU Symposium, ed. W.~J. {Jin}, I.~{Platais}, \&
  M.~A.~C. {Perryman}, 217--223

\bibitem[{{Lockwood}(2005)}]{lockwood05}
{Lockwood}, M. 2005, in Saas-Fee Advanced Course 34: The Sun, Solar Analogs and
  the Climate, ed. J.~D. {Haigh}, M.~{Lockwood}, M.~S. {Giampapa},
  I.~{R{\"u}edi}, M.~{G{\"u}del}, \& W.~{Schmutz}, 109--306

\bibitem[{Lockwood \& Frohlich(2007)}]{lockwood07}
Lockwood, M., \& Frohlich, C. 2007, Proceedings of the Royal Society A:
  Mathematical, Physical and Engineering Science, 463, 2447, 2447

\bibitem[{MacKay(2003)}]{mackay03}
MacKay, D. 2003, Information Theory, Inference and Learning Algorithms

\bibitem[{{Majaess} {et~al.}(2009){Majaess}, {Turner}, \& {Lane}}]{majaess09}
{Majaess}, D.~J., {Turner}, D.~G., \& {Lane}, D.~J. 2009, \mnras, 398, 263, 263

\bibitem[{Mart{\'\i} \& Ernst(2005)}]{marti05}
Mart{\'\i}, J., \& Ernst, G. 2005, Volcanoes And The Environment

\bibitem[{{Martos} {et~al.}(2004){Martos}, {Ya{\~n}ez}, {Hernandez}, {Moreno},
  \& {Pichardo}}]{martos04}
{Martos}, M., {Ya{\~n}ez}, M., {Hernandez}, X., {Moreno}, E., \& {Pichardo}, B.
  2004, Journal of Korean Astronomical Society, 37, 199, 199

\bibitem[{{Matese} {et~al.}(1995){Matese}, {Whitman}, {Innanen}, \&
  {Valtonen}}]{matese95}
{Matese}, J.~J., {Whitman}, P.~G., {Innanen}, K.~A., \& {Valtonen}, M.~J. 1995,
  \icarus, 116, 255, 255

\bibitem[{{Melott} {et~al.}(2012){Melott}, {Bambach}, {Petersen}, \&
  {McArthur}}]{melott12}
{Melott}, A.~L., {Bambach}, R.~K., {Petersen}, K.~D., \& {McArthur}, J.~M.
  2012, Journal of Geology, 120, 217, 217

\bibitem[{{Melott Adrian L.} \& {Bambach Richard K.}(2011)}]{melott10}
{Melott Adrian L.}, \& {Bambach Richard K.} 2011, Paleobiology, 37, 383, 383,
  doi: 10.1666/09055.1

\bibitem[{{Melott Adrian L.} \& {Thomas Brian C.}(2009)}]{melott08}
{Melott Adrian L.}, \& {Thomas Brian C.} 2009, Paleobiology, 35, 311, 311, doi:
  10.1666/0094-8373-35.3.311

\bibitem[{{Mishurov} \& {Acharova}(2011)}]{mishurov11}
{Mishurov}, Y.~N., \& {Acharova}, I.~A. 2011, \mnras, 412, 1771, 1771

\bibitem[{{Miyamoto} \& {Nagai}(1975)}]{miyamoto75}
{Miyamoto}, M., \& {Nagai}, R. 1975, \pasj, 27, 533, 533

\bibitem[{Muller \& MacDonald(2000)}]{muller00}
Muller, R., \& MacDonald, G. 2000, Ice Ages and Astronomical Causes: Data,
  Spectral Analysis, and Mechanisms, Springer-Praxis Books in Environmental
  Sciences

\bibitem[{{Muller}(1988)}]{muller88}
{Muller}, R.~A. 1988, LPI Contributions, 673, 127, 127

\bibitem[{{Overholt} {et~al.}(2009){Overholt}, {Melott}, \&
  {Pohl}}]{overholt09}
{Overholt}, A.~C., {Melott}, A.~L., \& {Pohl}, M. 2009, \apjl, 705, L101, L101

\bibitem[{Peters(2005)}]{peters05}
Peters, S.~E. 2005, Proceedings of the National Academy of Sciences of the
  United States of America, 102, 12326, 12326

\bibitem[{{Plummer}(1911)}]{plummer11}
{Plummer}, H.~C. 1911, \mnras, 71, 460, 460

\bibitem[{{Rampino} \& {Stothers}(1984)}]{rampino84}
{Rampino}, M.~R., \& {Stothers}, R.~B. 1984, \nat, 308, 709, 709

\bibitem[{{Rampino} \& {Stothers}(2000)}]{rampino00}
{Rampino}, M.~R., \& {Stothers}, R.~B. 2000, in Catastrophic Events and Mass
  Extinctions: Impacts and Beyond (University of Vienna, Austria: Lunar and
  Planetary Institute), 175

\bibitem[{{Raup} \& {Sepkoski}(1984)}]{raup84}
{Raup}, D.~M., \& {Sepkoski}, J.~J. 1984, Proceedings of the National Academy
  of Science, 81, 801, 801

\bibitem[{Raup(1972)}]{raup72}
Raup, D. M.~D. 1972, Science, 177, 1065–1071, 1065–1071

\bibitem[{{Rohde} \& {Muller}(2005)}]{rohde05}
{Rohde}, R.~A., \& {Muller}, R.~A. 2005, \nat, 434, 208, 208

\bibitem[{{Scalo} \& {Wheeler}(2002)}]{scalo02}
{Scalo}, J., \& {Wheeler}, J.~C. 2002, \apj, 566, 723, 723

\bibitem[{Sepkoski {et~al.}(2002)Sepkoski, Jablonski, \& Foote}]{sepkoski02}
Sepkoski, J., Jablonski, D., \& Foote, M. 2002, A Compendium of Fossil Marine
  Animal Genera, Bulletins of American paleontology

\bibitem[{{Sepkoski} \& {Raup}(1986)}]{sepkoski86}
{Sepkoski}, J.~J., \& {Raup}, D.~M. 1986, \nat, 321, 533, 533

\bibitem[{{Shaviv}(2003)}]{shaviv03}
{Shaviv}, N.~J. 2003, New Astronomy, 8, 39, 39

\bibitem[{{Shaviv}(2005)}]{shaviv05}
---. 2005, J. Geophys. Res., 110, A08105, A08105

\bibitem[{{Shoemaker}(1983)}]{shoemaker83}
{Shoemaker}, E.~M. 1983, Annual Review of Earth and Planetary Sciences, 11,
  461, 461

\bibitem[{{Sigurdsson}(1988)}]{sigurdsson88}
{Sigurdsson}, H. 1988, LPI Contributions, 673, 177, 177

\bibitem[{Sloan \& Wolfendale(2008)}]{sloan08}
Sloan, T., \& Wolfendale, A.~W. 2008, Environmental Research Letters, 3,
  024001, 024001

\bibitem[{{Stigler} \& {Wagner}(1987)}]{stigler87}
{Stigler}, S.~M., \& {Wagner}, M.~J. 1987, Science, 238, 940, 940

\bibitem[{{Thorsett}(1995)}]{thorsett95}
{Thorsett}, S.~E. 1995, \apjl, 444, L53, L53

\bibitem[{{Van Valen}(1973)}]{van-valen73}
{Van Valen}, L. 1973, Evolutionary Theory, 1, 1–30, 1–30

\bibitem[{{Vanhollebeke} {et~al.}(2009){Vanhollebeke}, {Groenewegen}, \&
  {Girardi}}]{vanhollebeke09}
{Vanhollebeke}, E., {Groenewegen}, M.~A.~T., \& {Girardi}, L. 2009, \aap, 498,
  95, 95

\bibitem[{{Wainscoat} {et~al.}(1992){Wainscoat}, {Cohen}, {Volk}, {Walker}, \&
  {Schwartz}}]{wainscoat92}
{Wainscoat}, R.~J., {Cohen}, M., {Volk}, K., {Walker}, H.~J., \& {Schwartz},
  D.~E. 1992, \apjs, 83, 111, 111

\bibitem[{Wignall(2001)}]{wignall01}
Wignall, P. 2001, Earth-Science Reviews, 53, 1 , 1

\bibitem[{Wignall {et~al.}(2009)Wignall, Sun, Bond, Izon, Newton, V�İdrine,
  Widdowson, Ali, Lai, Jiang, Cope, \& Bottrell}]{wignall09}
Wignall, P.~B., Sun, Y., Bond, D. P.~G., {et~al.} 2009, Science, 324, 1179,
  1179

\bibitem[{Winkler(1972)}]{winkler72}
Winkler, R. 1972, An Introduction to Bayesian Inference and Decision, Series in
  Quantitative Methods for Decision Making

\end{thebibliography}
\end{document}